\PassOptionsToPackage{unicode}{hyperref}
\PassOptionsToPackage{hyphens}{url}
\documentclass[
  10pt,
]{article}
\usepackage{xcolor}
\usepackage[margin=22mm]{geometry}
\usepackage{amsmath,amssymb}
\usepackage{iftex}
\ifPDFTeX
  \usepackage[T1]{fontenc}
  \usepackage[utf8]{inputenc}
  \usepackage{textcomp} 
  \DeclareUnicodeCharacter{2212}{\ensuremath{-}} 
\else 
  \usepackage{unicode-math} 
  \defaultfontfeatures{Scale=MatchLowercase}
  \defaultfontfeatures[\rmfamily]{Ligatures=TeX,Scale=1}
\fi
\usepackage{lmodern}
\ifPDFTeX\else
\fi
\IfFileExists{upquote.sty}{\usepackage{upquote}}{}
\IfFileExists{microtype.sty}{
  \usepackage[]{microtype}
  \UseMicrotypeSet[protrusion]{basicmath} 
}{}
\makeatletter
\@ifundefined{KOMAClassName}{
  \IfFileExists{parskip.sty}{%
    \usepackage{parskip}
  }{
    \setlength{\parindent}{0pt}
    \setlength{\parskip}{6pt plus 2pt minus 1pt}}
}{
  \KOMAoptions{parskip=half}}
\makeatother
\usepackage{longtable,booktabs,array}
\usepackage{caption}
\usepackage{calc} 
\usepackage{etoolbox}
\makeatletter
\patchcmd\longtable{\par}{\if@noskipsec\mbox{}\fi\par}{}{}
\makeatother
\IfFileExists{footnotehyper.sty}{\usepackage{footnotehyper}}{\usepackage{footnote}}
\makesavenoteenv{longtable}
\usepackage{graphicx}
\makeatletter
\newsavebox\pandoc@box
\newcommand*\pandocbounded[1]{
  \sbox\pandoc@box{#1}%
  \Gscale@div\@tempa{\textheight}{\dimexpr\ht\pandoc@box+\dp\pandoc@box\relax}%
  \Gscale@div\@tempb{\linewidth}{\wd\pandoc@box}%
  \ifdim\@tempb\p@<\@tempa\p@\let\@tempa\@tempb\fi
  \ifdim\@tempa\p@<\p@\scalebox{\@tempa}{\usebox\pandoc@box}%
  \else\usebox{\pandoc@box}%
  \fi%
}
\def\fps@figure{htbp}
\makeatother
\NewDocumentCommand\citeproctext{}{}
\NewDocumentCommand\citeproc{mm}{%
  \begingroup\def\citeproctext{#2}\cite{#1}\endgroup}
\makeatletter
 \let\@cite@ofmt\@firstofone
 \def\@biblabel#1{}
 \def\@cite#1#2{{#1\if@tempswa , #2\fi}}
\makeatother
\newlength{\cslhangindent}
\newlength{\csllabelwidth}
\newenvironment{CSLReferences}[2] 
 {\begin{list}{}{%
  \setlength{\itemindent}{0pt}
  \setlength{\leftmargin}{0pt}
  \setlength{\parsep}{0pt}
  \ifodd #1
   \setlength{\leftmargin}{\cslhangindent}
   \setlength{\itemindent}{-1\cslhangindent}
  \fi
  \setlength{\itemsep}{#2\baselineskip}}}
 {\end{list}}
\usepackage{calc}

\usepackage{float}
\usepackage{etoolbox}
\AtBeginEnvironment{longtable}{\small}
\usepackage{bookmark}
\IfFileExists{xurl.sty}{\usepackage{xurl}}{} 
\makeatletter
\@ifundefined{xmpquote}{\newcommand{\xmpquote}[1]{#1}}{}
\makeatother
\hypersetup{
  pdftitle={Pooling ETAS triggering parameters across subduction margins: forecast transfer{,} estimator stability and sequence-limited skill in Mexico},
  pdfauthor={Ricardo Mansilla Sánchez1; Ricardo Mansilla Corona2,3},
  pdfkeywords={\xmpquote{ETAS, earthquake forecasting, subduction zones,
information gain, model transfer, Mexico}},
  hidelinks,
  pdfcreator={LaTeX via pandoc}}

\title{Pooling ETAS triggering parameters across subduction margins:
forecast transfer, estimator stability and sequence-limited skill in
Mexico}
\author{Ricardo Mansilla Sánchez\textsuperscript{1} \and Ricardo
Mansilla Corona\textsuperscript{2,3}}
\date{12 September 2026}

\begin{document}
\maketitle
\begin{abstract}
Pooling earthquake-triggering parameters can reduce estimation
variability on margins with few well-recorded sequences, but its
forecast value must be distinguished from the stability obtained simply
by freezing a fitted law. We fit the epidemic-type aftershock sequence
(ETAS) model to the Mexican national catalog above magnitude 5.0, using
the Slab2 coverage mask as the observation and simulation domain. On a
4.7-year holdout the model gains 1.45 bits per earthquake over a
time-independent spatial Poisson reference, cluster-resampled interval
{[}−0.03, +2.75{]} bits, with two clusters carrying 90\% of the signed
gain. Using the Global Centroid Moment Tensor catalog we fit eight
disjoint subduction margins and a joint model on 3,210 training
earthquakes that shares six triggering parameters and retains a
background rate per margin; BIC prefers the joint model by 184 units,
although a parametric bootstrap rejects its exact shared specification.
Transferred into Mexico with the local background held fixed, six of
seven foreign triggering sets score within 0.07 nats per event of the
home fit. In leave-one-margin-out evaluation against local fits
restricted to the admissible domain, the excluded-pool law's
event-weighted paired gain is +0.0334 {[}+0.0002, +0.0670{]} nats per
event; its lower limit sits barely above zero and moves across zero with
the cluster-linking rule, so we read it neither as conclusive
superiority nor as a guarantee of noninferiority. In 28 overlapping
annual evaluations, local refitting minus a pooled law frozen before the
first origin has mean gain -2.07 nats per year, moving-block interval
{[}−6.92, +1.15{]}, while local productivity varies by a factor of 5.4
across origins. The coverage of the paired block intervals is
unmeasured, and the one measured analogue fell short of nominal; the
status of the paired median contrast depends on a Gutenberg--Richter
slope that is a convention rather than a measurement. These results
support pooling as a regularized alternative to local fitting; they stop
short of equivalence, of a universal physical law and of an optimal
update policy.
\end{abstract}

\textsuperscript{1}Arithmos Data Science.

\textsuperscript{2}Centro de Investigaciones Interdisciplinarias en
Ciencias y Humanidades (CEIICH), Universidad Nacional Autónoma de
México, Mexico City, Mexico.

\textsuperscript{3}Centro Peninsular en Humanidades y Ciencias Sociales
(CEPHCIS), Universidad Nacional Autónoma de México, Mérida, Yucatán,
Mexico.

Corresponding author: Ricardo Mansilla Sánchez (rick@arithmoslab.com).

\section{Introduction}\label{introduction}

Operational earthquake forecasting describes the rate of future
earthquakes, or a distribution for their number over a specified region
and interval; it does not identify an individual future rupture
(\citeproc{ref-Jordan2011}{Jordan et al. 2011};
\citeproc{ref-Schorlemmer2018}{Schorlemmer et al. 2018}). The
epidemic-type aftershock sequence model (ETAS) represents a background
process plus cascades in which every earthquake may trigger subsequent
events (\citeproc{ref-Ogata1988}{Ogata 1988},
\citeproc{ref-Ogata1998}{1998}). Its short-term advantage over a
stationary reference is often concentrated around large sequences
(\citeproc{ref-Helmstetter2006}{Helmstetter et al. 2006};
\citeproc{ref-Werner2011}{Werner et al. 2011}). Consequently, the number
of recorded earthquakes can greatly exceed the number of substantially
independent observations of forecast performance. Parameter estimation
and forecast evaluation need not have the same effective sample size.

On a margin with few well-recorded sequences, fitting a separate
triggering law can trade estimation variability against genuine regional
adaptation. Pooling offers an alternative: estimate common triggering
parameters across margins while retaining a local spatial background and
rate. Regional variation in ETAS parameters is documented
(\citeproc{ref-Chu2011}{Chu et al. 2011};
\citeproc{ref-Nandan2017}{Nandan et al. 2017}), but parameter
differences do not themselves establish useful out-of-sample score
differences. Catalog completeness, magnitude conventions, depth mixtures
and missing early aftershocks can also change the estimates
(\citeproc{ref-Sornette2005}{Sornette and Werner 2005};
\citeproc{ref-Hainzl2008}{Hainzl et al. 2008};
\citeproc{ref-Seif2017}{Seif et al. 2017};
\citeproc{ref-Mizrahi2021}{Mizrahi et al. 2021}). We treat shared
triggering parameters as a regularization hypothesis; physical identity
across margins is a separate claim.

Zhang, Werner and Goda (\citeproc{ref-Zhang2020}{Zhang et al. 2020})
studied ETAS variability across global subduction zones and proposed
representative parameter sets for aftershock hazard assessment. Their
parameter values are not directly interchangeable with ours, since ETAS
parameterizations differ in whether spatial scaling depends on magnitude
and in how kernels are normalized. Our contribution is a joint pooled
likelihood, transfer evaluated with controlled local backgrounds,
leave-one-margin-out scoring, and comparison of frozen, refitted and
partially pooled configurations over rolling origins.

We ask how much of the Mexican forecast gain is carried by a small
number of sequences, whether separately fitted triggering parameters
transfer when the recipient background is controlled, and whether local
refitting improves a specified forecast score enough to motivate
updating the triggering law. The primary operational comparison is the
gain accumulated during an annual evaluation window; mean and median
per-event gains are secondary summaries with a different meaning. That
designation was made after the rolling results existed, as a statement
of which quantity an operator should care about, and it was not
preregistered, so all three estimands are reported with their intervals.
We assign no numerical cost to forecast changes, and the results cannot
identify a uniquely optimal service policy.

\section{Data and observation
domains}\label{data-and-observation-domains}

\subsection{Mexican national catalog}\label{mexican-national-catalog}

The national catalog is supplied by the Servicio Sismológico Nacional
(SSN), Instituto de Geofísica, Universidad Nacional Autónoma de México
(\citeproc{ref-SSN2026}{Servicio Sismológico Nacional 2026};
\citeproc{ref-SSN2018}{Pérez-Campos et al. 2018}). The enclosing
rectangle is 107°W--88°W and 12°N--22°N. We retain events only where the
Central America segment of Slab2 has a defined interpolated interface
depth (\citeproc{ref-Hayes2018}{Hayes et al. 2018};
\citeproc{ref-Hayes2018data}{Hayes 2018}). In this paper Slab2 serves
only as a coverage mask. Interface, upper-plate and intraslab seismicity
coexist in this domain; the two large Mexican earthquakes of September
2017 were intraslab events with different rupture geometries
(\citeproc{ref-Melgar2018}{Melgar et al. 2018};
\citeproc{ref-Singh2018}{Singh et al. 2018}).

The observation domain is the intersection of the enclosing rectangle
with the valid Slab2 interpolation cells. A bilinear depth is defined
within cells whose four native grid corners are finite. We use that same
domain for catalog selection, background normalization, spatial kernel
integration and simulated offspring. Cells outside the mask are
unobserved, and the likelihood treats them that way. Off-domain
offspring are discarded before reproduction, so triggering through
outside events is omitted from both the likelihood and the simulations.

The adopted analysis threshold is SSN magnitude 5.0. The \(b\)-value
declines from 1.99 at an assumed cutoff of 4.0 to 1.020 with a standard
error of 0.035 above 5.0 for the whole rectangle, on 961 events; the
on-slab training subset that the model is actually fitted to gives 1.106
on 545 events, with an Aki standard error of 0.047
(\citeproc{ref-Aki1965}{Aki 1965}). A separate fixed value of 0.95 is
used wherever a magnitude distribution enters the triggering
calculations; it was fixed by convention, and the threshold-conversion
section states where it enters and what substituting either measured
value costs. The threshold selects a more homogeneous magnitude
population above 5.0 without certifying detection completeness in every
depth population (\citeproc{ref-Wiemer2000}{Wiemer and Wyss 2000}).
Inflated \(b\) at low cutoffs is not the classic signature of missing
small events, which usually depresses \(b\); mixed magnitude conventions
and population composition are alternative explanations, and the SSN
export does not record magnitude type. Regime-specific \(b\)-stability
thresholds extend to about 5.2, so mild residual incompleteness is
possible at the common cutoff. We also fit magnitude 4.5 as a
specification diagnostic, using an in-sample temporal residual test that
can reject a specification without saying why it fails.

The year 2000 is conditioning history, and the scored training window is
{[}2001-01-01, 2022-01-01). It contains 545 scored earthquakes of
magnitude 5.0 and above out of 560 training/history events. The holdout
is {[}2022-01-01, 2026-09-01), containing 101 events; its endpoint is
the day after the final dated official export used for this evaluation,
and events after it do not contribute to the score. Pre-2000 SSN
magnitude statistics vary substantially across eras, which motivates a
different catalog for longer histories.

The official SSN exports were generated on 2026-08-25 and 2026-08-31.
Overlapping exports are deduplicated by UTC timestamp, latitude and
longitude, retaining the later-generated row on a tie. The project also
accumulates a rolling recent-events feed, but all 101 evaluated holdout
events come from dated official exports. Provenance and review status
are separate: preliminary catalog status persists after April 2025
regardless of whether a row came from an official export. The evaluation
uses the final available catalog.

\begin{figure}
\centering
\includegraphics[width=1\linewidth,height=\textheight,keepaspectratio,alt={The Mexican study region and seismicity. Slab2 defines the observation mask; the enclosing rectangle is not itself the fitted observation domain. The cross-section and the 40-km depth split are descriptive and do not classify the triggering model into depth populations.}]{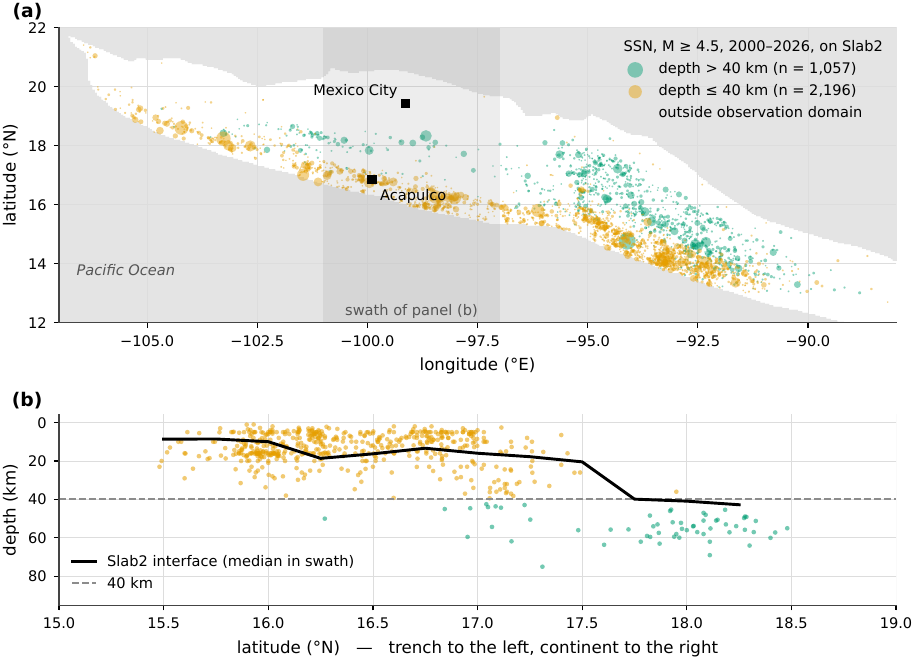}
\caption{The Mexican study region and seismicity. Slab2 defines the
observation mask; the enclosing rectangle is not itself the fitted
observation domain. The cross-section and the 40-km depth split are
descriptive and do not classify the triggering model into depth
populations.}
\end{figure}

\begin{figure}
\centering
\includegraphics[width=1\linewidth,height=\textheight,keepaspectratio,alt={Catalog homogeneity diagnostics for the Mexican region. SSN b-values vary strongly across early eras, whereas the GCMT series is more stable. This motivates the catalog boundary without identifying a unique cause for the SSN changes.}]{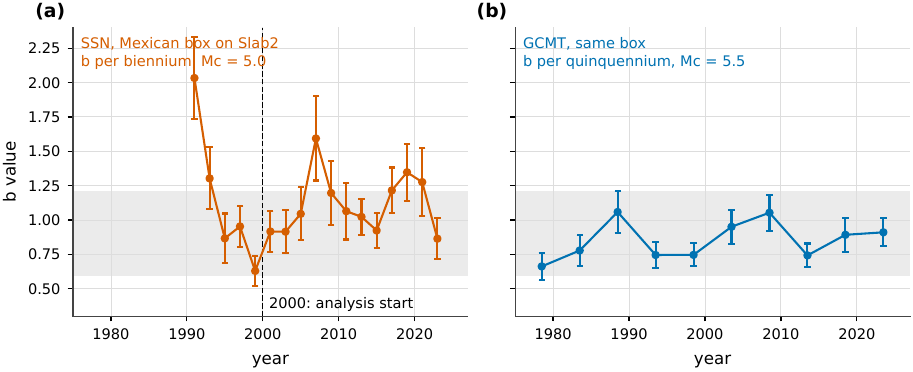}
\caption{Catalog homogeneity diagnostics for the Mexican region. SSN
\(b\)-values vary strongly across early eras, whereas the GCMT series is
more stable. This motivates the catalog boundary without identifying a
unique cause for the SSN changes.}
\end{figure}

\subsection{Global Centroid Moment Tensor
catalog}\label{global-centroid-moment-tensor-catalog}

The GCMT catalog contains 70,879 events from January 1976 through April
2026 (\citeproc{ref-Dziewonski1981}{Dziewonski et al. 1981};
\citeproc{ref-Ekstrom2012}{Ekström et al. 2012}). Magnitudes are derived
from scalar moments of centroid-moment-tensor inversions. We adopt Mw
5.5, supported by \(b\)-stability diagnostics across the selected
margins. A common magnitude definition does not imply an unchanged
observation process, since processing and detection of smaller events
changed over the catalog's history, and the diagnostics do not exclude
short-term incompleteness or every processing-era effect; the figure
below shows them margin by margin.

Events use centroid times and coordinates. The loader reads the
hypocentral date and time, adds the reported centroid time shift, and
computes Mw from scalar moment. A date-only parser would destroy sub-day
triggering structure while leaving counts and magnitude diagnostics
unchanged. Centroid shifts of seconds to tens of seconds are small
relative to the fitted onset scales of every fit in this paper.

Ten disjoint longitude--latitude boxes were specified. Eight clear the
operational minimum of 60 scored training events: Mexico, Central
America, Peru--Ecuador, Chile, Alaska--Aleutians, Japan,
Kamchatka--Kuriles, and Sumatra--Java. Cascadia and New Zealand do not.
Warm-up is 1976--1980, scored training is {[}1980-01-01, 2016-01-01),
and holdout is {[}2016-01-01, 2026-01-01). The eight margins contribute
3,210 scored training events and 909 holdout events; Mexico contributes
350 and 115, respectively. Scoring stops four months before the end of
the catalog.

Disjoint ownership prevents duplicate events in the joint likelihood;
cross-edge triggering is omitted, and disjointness says nothing about
physical independence across margins. Each GCMT box is its own
rectangular observation domain, whereas the Mexican SSN analysis uses
the Slab2 mask. Magnitude thresholds, domains and catalogs differ
between the two analyses, so their absolute per-event scores are not
interchangeable.

\begin{figure}
\centering
\includegraphics[width=1\linewidth,height=\textheight,keepaspectratio,alt={b-value stability by assumed magnitude threshold in the GCMT boxes. Bands show one standard error (Shi and Bolt 1982). The diagnostic motivates Mw 5.5; it does not certify unchanged completeness through time or immediately after large earthquakes.}]{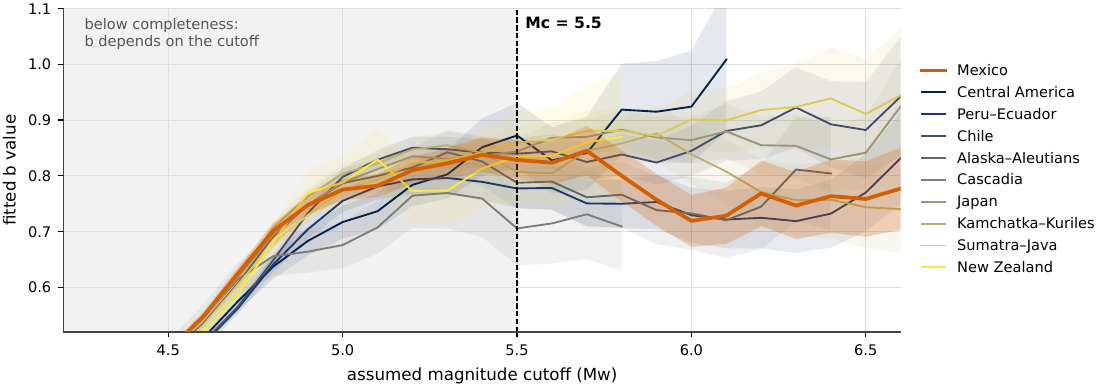}
\caption{\(b\)-value stability by assumed magnitude threshold in the
GCMT boxes. Bands show one standard error
(\citeproc{ref-ShiBolt1982}{Shi and Bolt 1982}). The diagnostic
motivates Mw 5.5; it does not certify unchanged completeness through
time or immediately after large earthquakes.}
\end{figure}

\section{Model, scoring and
uncertainty}\label{model-scoring-and-uncertainty}

\subsection{Intensity and
normalization}\label{intensity-and-normalization}

For an observation domain D, the conditional intensity is

\[\lambda(t,x,y)=\mathbf{1}_{D}(x,y)\left[\mu u(x,y)+\sum_{t_i<t}\nu_i g(t-t_i)f(r_i)\mathbf{1}\{t-t_i\le T,\ r_i\le R\}\right],\]

where

\[\nu_i=K\exp[\alpha(m_i-M_c)],\qquad g(s)=(p-1)c^{p-1}(s+c)^{-p},\]

\[f(r)=\frac{q-1}{\pi d^2}\left(1+\frac{r^2}{d^2}\right)^{-q}.\]

Times are measured in days. Spatial coordinates use an equirectangular
projection in kilometers, with longitude scaled at each region's
mid-latitude; distances and areas refer to this planar approximation.
The untruncated kernels integrate to one over positive time and the
plane. We truncate interactions at T=3,650 days and R=400 km without
renormalizing. \(K\) is thus the untruncated reference offspring count
of a parent at Mc; the expected offspring inside the fitted domain is
smaller. Ignoring domain boundaries, the retained fraction is

\[A_T=\left[1-\left(\frac{c}{T+c}\right)^{p-1}\right]\left[1-\left(1+\frac{R^2}{d^2}\right)^{1-q}\right].\]

A parent of magnitude m has cutoff-limited expected offspring
\(K\exp[\alpha(m-M_c)]A_T\), reduced further by domain boundaries.
Freeing \(c\), \(p\), \(d\) or \(q\) changes retained productivity even
if \(K\) and \(\alpha\) remain fixed.

For an unbounded Gutenberg--Richter magnitude distribution with
\(\beta=b\ln 10\) and \(\alpha<\beta\), the reference branching ratio is
\(n=K\beta/(\beta-\alpha)\). We also report \(n_T=nA_T\), a cutoff-only
reference that excludes domain-edge loss. Neither is the exact
reproduction criterion of the bounded masked process. We use \(n<1\) and
\(\alpha<\beta\) as a conservative admissibility guard on the
untruncated reference law. Ratios are labeled with their \(b\)-value.
Because \(\beta\) appears in both the numerator and the denominator,
\(n\) is far less sensitive to the assumed \(b\) than \(K\) alone is,
and the guard is correspondingly insensitive; the threshold-conversion
section quantifies this. Simulations use an explicitly upper-truncated
magnitude distribution; its average reproduction differs from this
unbounded reference.

The spatial background \(u\) is a variable-bandwidth Gaussian KDE of
training locations: fifth-neighbour bandwidths clipped to 5--75 km and a
uniform floor with weight 0.01. Both components are normalized on D.
Gaussian mass is integrated analytically over a rectangle or the
disjoint valid cells of the Slab2 mask; the uniform component uses the
same domain area. The KDE uses all training events, without iterative
background-probability weighting. Consequently it may absorb some
spatial clustering; both competing forecasts use the same estimated map,
but this construction can still affect estimated triggering parameters.

The point-process log-likelihood is the sum of event log intensities
minus the integrated intensity over the observation window. Warm-up
events trigger but do not enter the scored log sum. Temporal integrals
are analytic with the same cutoff as pair interactions. Spatial
integration uses the in-domain arc fraction on radial nodes, including
the actual 400-km endpoint, followed by quadrature in log radius. The
production grid has 280 log-spaced radial nodes over 0.001--5,000 km
before truncation and 256 angular samples. The spatial integral is
numerical; finer radial, angular and entropy-grid checks are reported
below.

Parameters are fitted by direct numerical maximization and by an
expectation-maximization algorithm over latent parent assignments
(\citeproc{ref-Veen2008}{Veen and Schoenberg 2008};
\citeproc{ref-Zhuang2002}{Zhuang et al. 2002}). Agreement between the
two checks the implementation and the reached solution; global
maximality is examined separately with multiple starting points,
observed curvature and profile-likelihood grids. Weakly identified
kernel parameters can have asymmetric or unclosed profiles, and Wald
standard errors are local summaries. The pooled model maximizes the sum
of per-domain likelihoods with six shared triggering parameters and one
\(\mu\) per margin.

\subsection{Scores and estimands}\label{scores-and-estimands}

The reference is a time-independent spatial Poisson process with the
same \(u\) and a constant rate estimated from the scored training
window. The information gain is

\[IG=\frac{\log L_E-\log L_P}{N}=\frac{\sum_i\log(\lambda_E/\lambda_P)-(\Lambda_E-\Lambda_P)}{N}.\]

This includes temporal and spatial triggering gains and the exposure
penalty (\citeproc{ref-Rhoades2011}{Rhoades et al. 2011};
\citeproc{ref-Schorlemmer2018}{Schorlemmer et al. 2018}). Exponentiating
the full per-event score gives a normalized likelihood ratio, which need
not equal a multiplicative improvement in event probability. One nat is
\(1/\ln 2\) bits.

For attribution, each observed event receives an equal share of the
compensator difference, so its contribution is
\(g_i=\log(\lambda_E/\lambda_P)-(\Lambda_E-\Lambda_P)/N\). Resampling
these contributions holds the \textbf{per-event share} fixed; total
allocated exposure changes with the resampled count. This bookkeeping
reconstructs the full-window score exactly; the allocation of exposure
to events is a convention.

Events are linked when separated by less than 30 days and 100 km.
Connected components, computed by union-find, are the cluster units.
Each of 20,000 bootstrap replicates draws the original number of
clusters with replacement and divides summed contributions by summed
event counts. Main intervals use seed 0 and percentile endpoints;
dropping the top k clusters uses seed k. Paired intervals resample
differences on the same events; marginal intervals of two candidates can
overlap or not regardless of the paired result, and their overlap is not
a test of the contrast. Throughout the paper a contrast is called
\emph{resolved} when its paired interval excludes zero and
\emph{unresolved} otherwise; the word describes the interval and carries
no claim about its coverage.

These cluster intervals are conditional sensitivity summaries whose
coverage is measured below under one process only, and two dominant
clusters strain them; the rule sweep in the results section shows how
far the headline interval moves with the partition. The linking
thresholds are a pragmatic dependence rule and nothing more: clusters so
defined need not be independent, and the rule is unrelated to ETAS
parent assignments. No noninferiority margin was predefined, so an
interval that includes zero is read as an unresolved contrast and never
as equivalence or as a bound on future losses. Point-process bootstrap
theory requires additional assumptions before it covers this
score-resampling construction (\citeproc{ref-Cavaliere2023}{Cavaliere et
al. 2023}; \citeproc{ref-Davison1997}{Davison and Hinkley 1997}).

Because the linking rule is a choice, three sets of cluster intervals
are also recomputed under eight alternative rules, all combinations of
15, 30 and 60 days with 50, 100 and 200 km, on the same per-event gains
(\texttt{sweep\_cluster.py}): the masked SSN holdout gain, the eight
leave-one-margin-out contrasts against the recipients' free own fits,
and the event-weighted aggregate of those recipients against their
admissible-restricted fits. Nothing is rescored in that sweep; only the
partition changes. The other cluster intervals in the paper, in the
transfer table, the correction ladder, the regime scores and the
memory-removal diagnostics, are not swept. The sweep's results are
reported next to the intervals they qualify.

The coverage of the cluster interval is measured by simulation under the
one process for which we can (\texttt{coverage\_cluster.py}): the fitted
SSN law generates 500 holdout windows conditioned on the real history,
each is scored and given the paper's cluster and iid intervals, and the
fraction covering the expected per-event gain of the process is
recorded. This is coverage under the fitted law, and it is reported
together with where the observed holdout falls among the simulated ones.

Rolling comparisons distinguish three estimands: the mean total gain in
an annual window, the equally weighted mean of annual per-event gains,
and the median annual per-event gain. Each can rank candidates
differently.

Every contrast between two candidates is \textbf{paired by origin}: we
form the difference at each of the 28 origins first and then summarize
that series of differences. A median contrast is therefore the median of
the 28 paired differences, which is not the difference between the two
candidates' medians reported in the summary table. The two quantities
can differ in magnitude and in sign, because the origins at which one
candidate leads need not be the origins that sit at either candidate's
own median. Both appear in this paper: the summary table gives each
candidate's own mean and median, and the contrast tables give paired
summaries. Only the paired form has an interval attached.

Non-circular moving-block bootstrap (\citeproc{ref-Kunsch1989}{Künsch
1989}) samples consecutive blocks of origins, concatenates enough blocks
to cover 28 origins, and trims to 28. We report all block lengths 1--8,
20,000 replicates and seed 0. Blocks acknowledge dependence without
demonstrating calibrated coverage, so the coverage is measured
(\texttt{coverage\_rolling.py}): 200 synthetic catalogs from 2000 to
2026 are generated by the frozen pooled law C with the real training
background, each is replayed through the same 28 origins for C and for
the frozen local law D, and the fraction of block intervals covering the
expectation of the statistic under that process is recorded at every
block length. The primary contrast A−C, which refits seven parameters at
every origin, is about thirty times the cost and is not simulated; D−C
uses the same evaluation windows and resampling scheme but a different
estimation procedure, since its local law is estimated once per catalog
rather than at every origin; C alone shares only the dependence
structure. Wilcoxon tests and the AR(1) effective-size expression are
descriptive because annual windows overlap and training windows are
nested (\citeproc{ref-Wilcoxon1945}{Wilcoxon 1945}).

The temporal residual KS diagnostic is in sample and is not corrected
for fitted parameters; it can flag a rejected temporal specification,
and it says nothing about the spatial or magnitude model. A realized
compensator integrates the conditional intensity along the observed
path. A forecast count issued at the beginning of that path is a
different quantity, obtained by simulating background events, offspring
of conditioning history and within-window cascades
(\citeproc{ref-Harte2017}{Harte 2017}). Simulations use the same domain
and cutoffs, reject pre-window generated offspring of conditioning
parents before reproduction, and draw continuous upper-truncated
Gutenberg--Richter magnitudes. Generation caps raise an error if reached
with a live frontier.

BIC uses the scored training-event count as its sample-size penalty.
With a data-estimated KDE and possible misspecification it compares
penalized fits and yields no calibrated posterior model probability
(\citeproc{ref-Schwarz1978}{Schwarz 1978};
\citeproc{ref-White1982}{White 1982}). The parametric LR bootstrap
re-estimates the background and refits free and pooled models; its
recovery shifts include nuisance-background estimation effects as well
as finite-sample parameter estimation. Monte Carlo \(p\)-values use the
finite-replicate adjustment (k+1)/(B+1), with exceedance counts reported
(\citeproc{ref-Phipson2010}{Phipson and Smyth 2010}).

\section{Skill and sequence concentration on the Mexican
holdout}\label{skill-and-sequence-concentration-on-the-mexican-holdout}

\begin{longtable}[]{@{}lll@{}}
\caption{SSN fit on the masked observation domain, magnitude 5.0; 545
scored training events. \(K\) is the untruncated reference
amplitude.}\tabularnewline
\toprule\noalign{}
Parameter & Estimate & Local standard error \\
\midrule\noalign{}
\endfirsthead
\toprule\noalign{}
Parameter & Estimate & Local standard error \\
\midrule\noalign{}
\endhead
\bottomrule\noalign{}
\endlastfoot
\(\mu\) & 0.050212 & 0.00304 \\
\(K\) & 0.21507 & 0.0808 \\
\(\alpha\) & 1.355 & 0.122 \\
\(c\) & 0.034188 & 0.0176 \\
\(p\) & 1.1085 & 0.0597 \\
\(d\) & 10.565 & 3.03 \\
\(q\) & 1.4065 & 0.152 \\
\end{longtable}

At the lower diagnostic threshold of 4.5, EM reaches \(p=1\) with
\(K=4.79\times10^{13}\) and a reference branching ratio of
\(7.4\times10^{13}\). This boundary solution is inadmissible under the
reference-branching guard and has undefined local standard errors. The
direct optimiser finds a different, interior solution. The
lower-threshold run thus meets the EM stopping criterion without
yielding a well-determined admissible model, and a residual-test
\(p\)-value alone cannot rescue it.

The reference branching ratio is 0.460 at the training subset's
\(b=1.106\), and 0.565 at the conventional \(b=0.95\). The corresponding
cutoff-only ratios are 0.312 and 0.383; the irregular domain further
reduces offspring mass. The in-sample temporal KS \(p\)-value is 0.766.
Numerical curvature and profiles are reported in the identifiability
section.

On the 101-event holdout the gain is \textbf{1.45 bits/event}, or 1.003
nats/event. The cluster interval is {[}−0.019, +1.908{]} nats, against
the independent-event bootstrap's {[}+0.535, +1.510{]}. The latter is
shown only to illustrate the effect of the dependence assumption. The
model has a positive allocated contribution on 27.7\% of events, a
fraction that says where the gain lives and nothing about skill.

The two largest contributing connected components provide 90.0\% of the
signed total gain. Removing the highest-contributing clusters without
refitting gives mean gains of 1.003, 0.484, 0.127, 0.013 nats/event for
zero through three removals. This is attribution conditional on the
observed history; the forecast's score in a history without those
sequences is a different quantity, and so is the effective sample size,
which the count of dominant clusters only bounds.

The interval's status depends on the linking rule. Under the nine rules
the same 101 gains partition into 34 to 79 clusters, the two largest
carry 72\% to 90\% of the signed total, and the lower limit runs from
−0.020 to +0.123 nats/event while the upper limit stays between +1.85
and +2.06. The interval excludes zero under six of the nine rules and
includes it under the main rule, which is among the three that do not.
We keep the main rule because it was fixed before this sweep, and we
report the headline interval as rule-dependent: the sweep shows that no
linking rule in this range moves the lower limit further than 0.12
nats/event from zero in either direction, and it leaves the gain
unresolved.

\begin{longtable}[]{@{}
  >{\raggedright\arraybackslash}p{(\linewidth - 8\tabcolsep) * \real{0.2000}}
  >{\raggedright\arraybackslash}p{(\linewidth - 8\tabcolsep) * \real{0.2000}}
  >{\raggedright\arraybackslash}p{(\linewidth - 8\tabcolsep) * \real{0.2000}}
  >{\raggedright\arraybackslash}p{(\linewidth - 8\tabcolsep) * \real{0.2000}}
  >{\raggedright\arraybackslash}p{(\linewidth - 8\tabcolsep) * \real{0.2000}}@{}}
\caption{The masked SSN holdout gain under nine cluster-linking rules.
Same per-event gains, same 20,000 replicates and seed; only the
partition changes.}\tabularnewline
\toprule\noalign{}
\begin{minipage}[b]{\linewidth}\raggedright
Window (d)
\end{minipage} & \begin{minipage}[b]{\linewidth}\raggedright
Link (km)
\end{minipage} & \begin{minipage}[b]{\linewidth}\raggedright
Clusters
\end{minipage} & \begin{minipage}[b]{\linewidth}\raggedright
Top-2 share
\end{minipage} & \begin{minipage}[b]{\linewidth}\raggedright
Interval, nats/event
\end{minipage} \\
\midrule\noalign{}
\endfirsthead
\toprule\noalign{}
\begin{minipage}[b]{\linewidth}\raggedright
Window (d)
\end{minipage} & \begin{minipage}[b]{\linewidth}\raggedright
Link (km)
\end{minipage} & \begin{minipage}[b]{\linewidth}\raggedright
Clusters
\end{minipage} & \begin{minipage}[b]{\linewidth}\raggedright
Top-2 share
\end{minipage} & \begin{minipage}[b]{\linewidth}\raggedright
Interval, nats/event
\end{minipage} \\
\midrule\noalign{}
\endhead
\bottomrule\noalign{}
\endlastfoot
15 & 50 & 79 & 72\% & {[}+0.123, +1.854{]} \\
15 & 100 & 76 & 85\% & {[}+0.028, +1.896{]} \\
15 & 200 & 66 & 85\% & {[}+0.034, +1.899{]} \\
30 & 50 & 76 & 76\% & {[}+0.084, +1.861{]} \\
30 & 100 & 71 & 90\% & {[}−0.019, +1.908{]} (main) \\
30 & 200 & 52 & 90\% & {[}−0.020, +1.946{]} \\
60 & 50 & 68 & 76\% & {[}+0.078, +1.875{]} \\
60 & 100 & 56 & 90\% & {[}−0.007, +1.924{]} \\
60 & 200 & 34 & 90\% & {[}+0.003, +2.064{]} \\
\end{longtable}

Under the fitted law the cluster interval covers the process's expected
gain in 89.2\% of 500 simulated holdouts against a nominal 95\%, and the
iid-event interval in 79.2\%; the 10-point gap is the measured cost of
resampling events instead of sequences. The cluster interval's misses
are almost all on one side, 10.4\% of intervals falling entirely below
the target against 0.4\% above, which is the percentile bootstrap on a
right-skewed mean: a holdout that draws no large sequence produces a low
gain and an interval that does not reach the expectation. The observed
holdout is not a typical draw from that law. Its two largest clusters
carry 90\% of the gain, more than in 99.6\% of the simulated holdouts,
where the typical share is 47\%; its interval is wider than 95\% of the
simulated ones; and only 1.6\% of simulated intervals include zero. The
measured 89\% therefore applies to a process less concentrated than the
one observed, and it bounds how far the interval can be trusted: in the
regime of the observed holdout coverage is not established, and the
fitted law's failure to generate that concentration is itself a misfit
signal, consistent with the kernel-shape deficit of the correction
ladder.

\begin{figure}
\centering
\includegraphics[width=1\linewidth,height=\textheight,keepaspectratio,alt={Per-event gains, cumulative attribution and removal of the highest-contributing clusters on the masked SSN domain. The drop ladder is an attribution analysis without refitting.}]{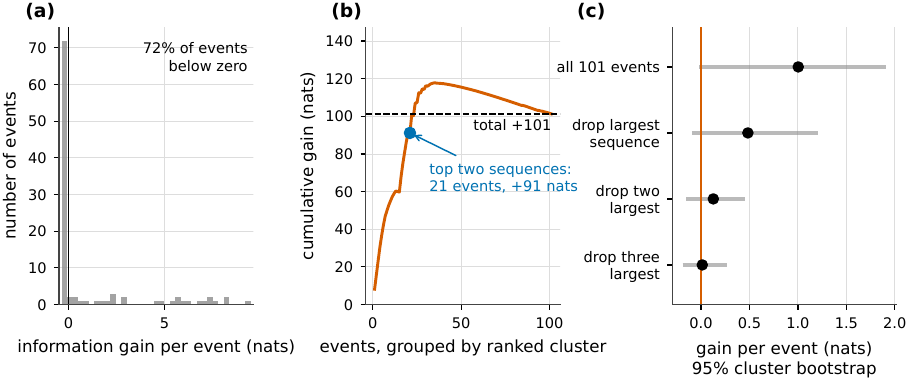}
\caption{Per-event gains, cumulative attribution and removal of the
highest-contributing clusters on the masked SSN domain. The drop ladder
is an attribution analysis without refitting.}
\end{figure}

Integrated over the holdout window and domain, the Poisson reference
gives an expected count of 121.08 events and the ETAS compensator along
the realized history 121.30, against 101 observed. From 1,000 forward
simulations conditioned on training history, the ETAS count mean is
128.88, median 128, and central 95\% range {[}98, 166{]}. The
variance-to-mean ratio is 2.51; empirical probabilities of counts at
most and at least 101 are 0.051 and 0.957. Magnitudes in this experiment
are continuously truncated at 8.4, and drawn with the fixed \(b\)
discussed in the threshold-conversion section, which sets the simulated
magnitude distribution and hence cascade sizes.

The observation of 101 events sits in the lower tail of that predictive
distribution: the one-sided probability of a count at most 101 is 0.051,
the equal-tailed two-sided probability is 0.10, and 101 lies inside the
central 95\% range. The simulated mean exceeds the observed count, and
we read the excess as a calibration signal, short of a rejection. The
distribution is conditional on one fitted parameter set, so it
understates the spread that parameter uncertainty would add; a single
holdout window is one draw, and the same window is used for every other
result in this section; and the experiment does not attribute the
shortfall. The realized compensator of 121.30 integrates the fitted
intensity along the observed history, and it decomposes exactly into a
background part, \(\mu\) times the window length, of 85.56 events and a
triggered part of 35.74, so triggering supplies 29\% of the ETAS
expectation along that history. Its closeness to the Poisson reference's
121.08 is therefore not evidence that the observed sequence triggered
little: the two models have different background rates, 0.0502 against
0.0711 events per day, and the triggered part fills the gap. The
simulated mean of 128.88 exceeds the compensator because the compensator
conditions on the realized history, whose 101 events supplied fewer
parents than the law expects on average, whereas the simulations
generate their own parents and average over histories; the difference
measures that conditioning, not the amount of triggering in the observed
data. Neither quantity attributes the low observed count to the
background or to the triggering component, and separating the two would
require intervening on the model's components. A prospective calibration
check on repeated windows would settle it.

The separate per-origin count diagnostic uses a 365.25-day window and
the training-rate \textbf{Poisson} reference. Rejections of that
reference cannot establish miscalibration of ETAS forecast counts or
uniquely attribute variation to a nonstationary background. Branching
itself produces overdispersion, so those diagnostics are retained as
descriptive checks on the constant-rate comparator; they do not diagnose
the background independently of the triggering law.

\subsection{Regime conditioning}\label{regime-conditioning}

Define \(S(t)=H_{\rm bg}-H(\lambda_t)\) using the normalized spatial
intensity on the masked forecast grid. The same 400-km and 3,650-day
cutoffs are enforced. Grid points are cell midpoints inside the
observation domain. S may be negative when the forecast is more
spatially diffuse than the background. The field is evaluated from
events more than one second earlier, avoiding accidental inclusion of an
event at its own timestamp. S is a deterministic summary of the fitted
field and inherits its dependence on the domain, background
construction, fitted parameters and spatial discretization.

We report both an allocated decomposition and a regime-exposure score.
The former splits the event contributions above by the flag. The latter
integrates the intensity difference only while the flag is active and
divides by the number of events in that regime. Segment integrals are
analytic in time conditional on the numerically computed spatial mass.
Threshold crossings are located numerically and refined. The regime
bootstrap holds the regime's total exposure fixed while resampling its
event clusters; it therefore differs from the bootstrap of allocated
per-event contributions.

\begin{longtable}[]{@{}
  >{\raggedright\arraybackslash}p{(\linewidth - 10\tabcolsep) * \real{0.1667}}
  >{\raggedright\arraybackslash}p{(\linewidth - 10\tabcolsep) * \real{0.1667}}
  >{\raggedright\arraybackslash}p{(\linewidth - 10\tabcolsep) * \real{0.1667}}
  >{\raggedright\arraybackslash}p{(\linewidth - 10\tabcolsep) * \real{0.1667}}
  >{\raggedright\arraybackslash}p{(\linewidth - 10\tabcolsep) * \real{0.1667}}
  >{\raggedright\arraybackslash}p{(\linewidth - 10\tabcolsep) * \real{0.1667}}@{}}
\caption{Exposure-restricted regime scores, bits/event, with conditional
cluster intervals. Allocated decompositions and convergence checks are
supplied separately.}\tabularnewline
\toprule\noalign{}
\begin{minipage}[b]{\linewidth}\raggedright
Threshold
\end{minipage} & \begin{minipage}[b]{\linewidth}\raggedright
Duty cycle
\end{minipage} & \begin{minipage}[b]{\linewidth}\raggedright
Sharp events
\end{minipage} & \begin{minipage}[b]{\linewidth}\raggedright
Sharp clusters
\end{minipage} & \begin{minipage}[b]{\linewidth}\raggedright
Sharp score
\end{minipage} & \begin{minipage}[b]{\linewidth}\raggedright
Quiet score
\end{minipage} \\
\midrule\noalign{}
\endfirsthead
\toprule\noalign{}
\begin{minipage}[b]{\linewidth}\raggedright
Threshold
\end{minipage} & \begin{minipage}[b]{\linewidth}\raggedright
Duty cycle
\end{minipage} & \begin{minipage}[b]{\linewidth}\raggedright
Sharp events
\end{minipage} & \begin{minipage}[b]{\linewidth}\raggedright
Sharp clusters
\end{minipage} & \begin{minipage}[b]{\linewidth}\raggedright
Sharp score
\end{minipage} & \begin{minipage}[b]{\linewidth}\raggedright
Quiet score
\end{minipage} \\
\midrule\noalign{}
\endhead
\bottomrule\noalign{}
\endlastfoot
\(S \ge 0.05\) & 29.20\% & 44 & 26 & 3.20 {[}−0.21, +4.98{]} & 0.10
{[}+0.03, +0.18{]} \\
\(S \ge 0.25\) & 13.55\% & 28 & 12 & 4.97 {[}−0.02, +6.61{]} & 0.10
{[}+0.01, +0.20{]} \\
\(S \ge 1\) & 4.24\% & 20 & 7 & 6.48 {[}+1.47, +7.83{]} & 0.20 {[}+0.01,
+0.45{]} \\
\(S \ge 2\) & 1.30\% & 17 & 6 & 7.65 {[}+3.41, +8.66{]} & 0.19 {[}−0.07,
+0.53{]} \\
\end{longtable}

The thresholds were chosen after inspecting the data. The flag is
computable before an event, but this ladder is exploratory and does not
validate a prospectively selected alarm policy. High sharpening can
identify a concentrated sequence without adding information beyond
triggering recency. Quiet time is not equivalent to the absence of a
sequence: a tail can continue after the flag drops. Positive quiet-time
scores can arise when the forecast correctly assigns fewer events than
the Poisson reference, even if its intensity at most isolated events is
lower.

\begin{figure}
\centering
\includegraphics[width=1\linewidth,height=\textheight,keepaspectratio,alt={Allocated regime decomposition and event-level sharpening. Exposure-restricted scores are reported in the accompanying table; the two constructions measure different things. Duty cycles use refined temporal crossings. Panel (b) preserves negative S values on a symmetric-log axis. Thresholds are exploratory.}]{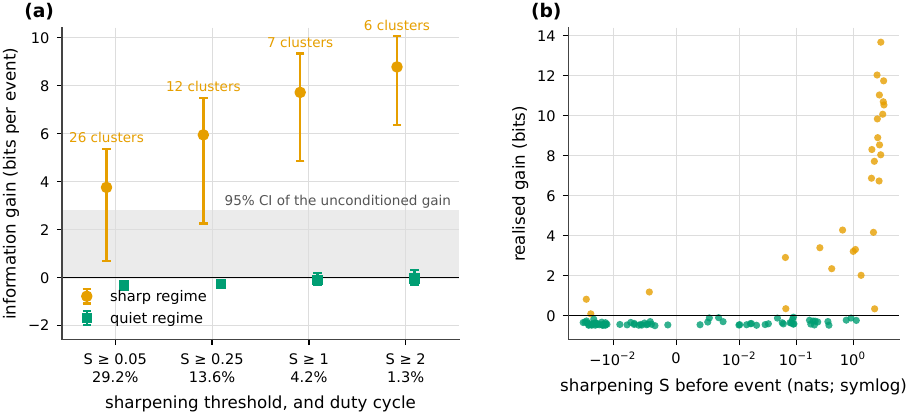}
\caption{Allocated regime decomposition and event-level sharpening.
Exposure-restricted scores are reported in the accompanying table; the
two constructions measure different things. Duty cycles use refined
temporal crossings. Panel (b) preserves negative S values on a
symmetric-log axis. Thresholds are exploratory.}
\end{figure}

\section{Transfer of triggering parameters between
margins}\label{transfer-of-triggering-parameters-between-margins}

Each donor contributes only its six triggering parameters. Mexico's ETAS
background rate \(\mu\), spatial density \(u\), training-rate Poisson
reference, conditioning history and holdout events are held at their
home values across rows. Transferring the donor \(\mu\) as well would
also transfer that margin's rate level, and a donor could then score
badly in Mexico because of its background rate alone; holding the rate
fixed removes that confound.

Holding \(\mu\) fixed is not, however, a neutral control, and we do not
claim it is. Mexico's \(\mu\) was estimated jointly with Mexico's own
triggering parameters, so it is the rate that makes the total expected
count correct \textbf{under the home law}. A donor law with different
retained productivity puts a different fraction of the total rate into
triggered events, and with \(\mu\) pinned that difference is not
reabsorbed: the exposure term of the score then varies with the donor's
\(K\) and \(\alpha\), not only with its kernel shapes. The size of the
effect is not negligible. Evaluated at Mexico's own \(b\) of 0.81, the
cutoff-limited branching ratios of six of the seven donor laws span 0.52
to 1.21 times the home value; the remaining donor has \(\alpha\) above
\(\beta\) at that \(b\) and so has no finite reference ratio. A donor is
thus scored partly on a rate calibration it never chose, and a factor of
two in retained productivity is a large perturbation of the exposure
term.

The experiment that isolates triggering is the opposite one: transfer
the six parameters and re-estimate \(\mu\) on the recipient's own
training data. That is the leave-one-margin-out design below, and it is
the version we rely on for the transfer claim. Where both can be
computed they agree. The pooled law scored against Mexico's home fit
differs by +0.023 nats/event with \(\mu\) adapted to the pooled law, and
the leave-one-margin-out row for Mexico, whose triggering parameters
never saw Mexico, gives +0.022. The fixed-\(\mu\) table is kept because
it answers a different and also interpretable question, namely how a
foreign kernel performs on a rate field it did not calibrate; the
leave-one-margin-out design is the stronger of the two.

No parameter is selected using the Mexican holdout in either experiment.
Both are scored on the same 115 Mexican GCMT events.

\begin{longtable}[]{@{}
  >{\raggedright\arraybackslash}p{(\linewidth - 8\tabcolsep) * \real{0.2000}}
  >{\raggedright\arraybackslash}p{(\linewidth - 8\tabcolsep) * \real{0.2000}}
  >{\raggedright\arraybackslash}p{(\linewidth - 8\tabcolsep) * \real{0.2000}}
  >{\raggedright\arraybackslash}p{(\linewidth - 8\tabcolsep) * \real{0.2000}}
  >{\raggedright\arraybackslash}p{(\linewidth - 8\tabcolsep) * \real{0.2000}}@{}}
\caption{Triggering-only transfer into Mexico; nats/event. The Mexican
ETAS background rate is fixed in every row.}\tabularnewline
\toprule\noalign{}
\begin{minipage}[b]{\linewidth}\raggedright
Donor
\end{minipage} & \begin{minipage}[b]{\linewidth}\raggedright
IG
\end{minipage} & \begin{minipage}[b]{\linewidth}\raggedright
Cluster interval
\end{minipage} & \begin{minipage}[b]{\linewidth}\raggedright
Foreign − home
\end{minipage} & \begin{minipage}[b]{\linewidth}\raggedright
Paired interval
\end{minipage} \\
\midrule\noalign{}
\endfirsthead
\toprule\noalign{}
\begin{minipage}[b]{\linewidth}\raggedright
Donor
\end{minipage} & \begin{minipage}[b]{\linewidth}\raggedright
IG
\end{minipage} & \begin{minipage}[b]{\linewidth}\raggedright
Cluster interval
\end{minipage} & \begin{minipage}[b]{\linewidth}\raggedright
Foreign − home
\end{minipage} & \begin{minipage}[b]{\linewidth}\raggedright
Paired interval
\end{minipage} \\
\midrule\noalign{}
\endhead
\bottomrule\noalign{}
\endlastfoot
Sumatra--Java & +0.732 & {[}+0.078, +1.425{]} & +0.034 & {[}−0.040,
+0.125{]} \\
Chile & +0.724 & {[}+0.061, +1.423{]} & +0.027 & {[}−0.058, +0.132{]} \\
Mexico & +0.698 & {[}+0.046, +1.419{]} & --- & --- \\
Japan & +0.696 & {[}+0.014, +1.420{]} & −0.002 & {[}−0.080, +0.090{]} \\
Kamchatka--Kuriles & +0.691 & {[}+0.067, +1.334{]} & −0.007 & {[}−0.138,
+0.138{]} \\
Peru--Ecuador & +0.646 & {[}+0.039, +1.302{]} & −0.052 & {[}−0.134,
+0.023{]} \\
Central America & +0.646 & {[}−0.112, +1.516{]} & −0.052 & {[}−0.190,
+0.111{]} \\
Alaska--Aleutians & +0.580 & {[}+0.062, +1.098{]} & −0.118 & {[}−0.367,
+0.108{]} \\
\end{longtable}

Six of the seven foreign point estimates lie within 0.07 nats/event of
the home fit, and no paired interval excludes zero. The foreign median
score is 0.691 against the home score 0.698. These are transfer results,
and equivalence would be a stronger claim than they support. With the
background rate controlled, the two smallest donor samples show no clear
transfer deficit; their small training samples and weak parameter
determination remain estimation limitations without being identified
here as the cause of a resolved deficit.

\begin{figure}
\centering
\includegraphics[width=0.78\linewidth,height=\textheight,keepaspectratio,alt={Controlled transfer into Mexico. Foreign rows retain Mexico's background rate and density; bars show marginal cluster intervals. Paired differences are in the table. The pooled and early-Mexico candidates use the procedures specified in the next section.}]{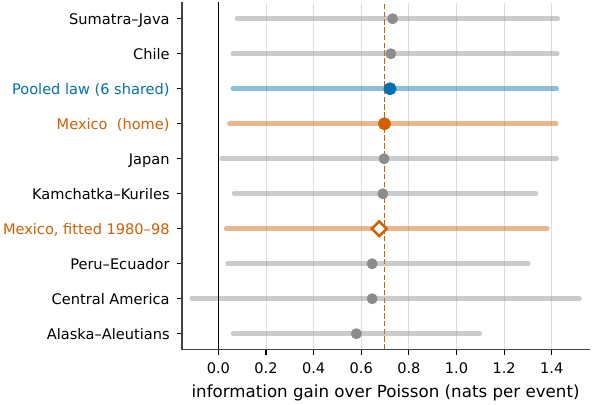}
\caption{Controlled transfer into Mexico. Foreign rows retain Mexico's
background rate and density; bars show marginal cluster intervals.
Paired differences are in the table. The pooled and early-Mexico
candidates use the procedures specified in the next section.}
\end{figure}

At their own estimated \(b\)-values, two free regional fits fail the
reference admissibility guard. They are retained as unconstrained
diagnostic fits in the LR comparison. A parameter set's admissibility at
its donor \(b\)-value is distinct from its finite-window Mexican score.

\section{A pooled law and held-out
recipients}\label{a-pooled-law-and-held-out-recipients}

The pooled model shares \(K\), \(\alpha\), \(c\), \(p\), \(d\) and \(q\)
across the eight margins and estimates a separate \(\mu\) for each. Each
regional background density is constructed from that region's training
catalog. Ignoring cross-domain triggering, the joint log-likelihood is
the sum of the regional conditional log-likelihoods. The free model has
56 parameters and the pooled model 14; the nominal difference is 42
degrees of freedom.

\begin{longtable}[]{@{}llll@{}}
\caption{Pooled and free GCMT fits on 3,210 scored training
events.}\tabularnewline
\toprule\noalign{}
Model & Parameters & Training log L & BIC \\
\midrule\noalign{}
\endfirsthead
\toprule\noalign{}
Model & Parameters & Training log L & BIC \\
\midrule\noalign{}
\endhead
\bottomrule\noalign{}
\endlastfoot
Shared triggering; local \(\mu\) & 14 & -54,077.52 & 108,268.08 \\
Seven parameters per region & 56 & -54,000.07 & 108,452.29 \\
\end{longtable}

The LR statistic is 154.90, with nominal \(\chi^2\)
\(p=7.7\times10^{-15}\); BIC prefers pooling by 184.2. Constraining the
inadmissible free fits to \(\alpha<\beta\) and \(n<1\) costs 0.793
log-likelihood units. The constrained statistic is 153.31; boundary
optima and a data-estimated background prevent interpreting the nominal
\(\chi^2\) result as an exact significance calculation.

In 200 parametric-bootstrap replicates from the fitted shared
specification, the unconstrained LR has mean 41.32, standard deviation
8.27, 95th percentile 53.93, and maximum 73.54. The observed
unconstrained statistic is exceeded by 0 replicates, giving adjusted
Monte Carlo \(p=0.0050\). With zero exceedances, the one-sided 95\%
binomial upper limit on the underlying exceedance probability is
approximately 0.015. This simulation calibrates the
\textbf{unconstrained} statistic.

The same replicates measure how well the pooled estimator recovers its
own generating law, and the shifts are not negligible. Against a truth
of \(K=0.149\), \(c=0.141\) and \(p=1.206\), the median refitted values
are 0.112, 0.205 and 1.337: productivity is recovered about 25\% low
while the onset time and decay exponent come back high. \(\alpha\),
\(d\) and \(q\) are recovered within about 5\%. The displaced
coordinates are the same near-flat direction that the identifiability
diagnostics single out, so the estimator moves along the likelihood
ridge.

The rejection concerns the particular shared statistical specification.
It does not separate physical regional variation from omitted depth
structure, catalog effects, background estimation or other model
misspecification. Re-estimating a KDE from all simulated events can also
contribute to these recovery shifts, so they are not attributed uniquely
to ordinary finite-sample MLE bias. A known-background control would be
needed to isolate that mechanism. The shifts are biases of the estimator
under its own null, so they qualify the interpretation of the pooled
coordinates and leave the observed holdout scores unchanged.

\begin{longtable}[]{@{}llllll@{}}
\caption{Shared law at the native GCMT threshold Mw 5.5. All \(K\)
values use the untruncated kernel normalization.}\tabularnewline
\toprule\noalign{}
\(K\) & \(\alpha\) & \(c\) (days) & \(p\) & \(d\) (km) & \(q\) \\
\midrule\noalign{}
\endfirsthead
\toprule\noalign{}
\(K\) & \(\alpha\) & \(c\) (days) & \(p\) & \(d\) (km) & \(q\) \\
\midrule\noalign{}
\endhead
\bottomrule\noalign{}
\endlastfoot
0.14902 & 1.2652 & 0.14099 & 1.2062 & 19.187 & 1.6021 \\
\end{longtable}

On the Mexican GCMT holdout, the pooled law has gain 0.721 and paired
difference from the home fit +0.023 {[}−0.050, +0.114{]} nats/event. The
fit made using Mexico's 1980--1998 training period has difference −0.022
{[}−0.045, +0.001{]}. The latter evaluates old \textbf{parameters} with
the conditioning history available before each scored event, so it is a
test of parameter aging with a current history.

\begin{figure}
\centering
\includegraphics[width=1\linewidth,height=\textheight,keepaspectratio,alt={Regional free parameters against the pooled fit. Open markers identify fits failing the reference admissibility guard. Differences in fitted coordinates do not by themselves establish differences in out-of-sample utility.}]{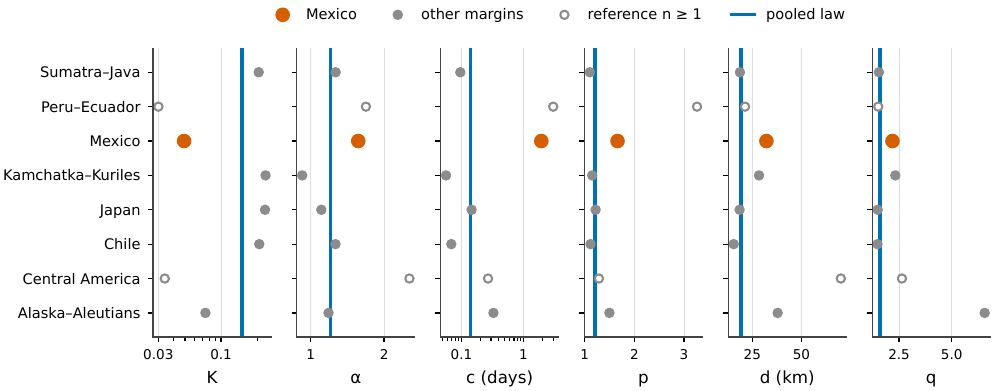}
\caption{Regional free parameters against the pooled fit. Open markers
identify fits failing the reference admissibility guard. Differences in
fitted coordinates do not by themselves establish differences in
out-of-sample utility.}
\end{figure}

\subsection{Leave-one-margin-out
evaluation}\label{leave-one-margin-out-evaluation}

For each recipient we refit the pooled law on the other seven margins,
estimate only \(\mu\) on the recipient's training data, and score the
recipient's holdout. The own-fit comparator uses that recipient's free
regional fit. This tests a transferred pooled law whose triggering
parameters were estimated without the recipient, while retaining local
rate adaptation.

\begin{longtable}[]{@{}
  >{\raggedright\arraybackslash}p{(\linewidth - 10\tabcolsep) * \real{0.1667}}
  >{\raggedright\arraybackslash}p{(\linewidth - 10\tabcolsep) * \real{0.1667}}
  >{\raggedright\arraybackslash}p{(\linewidth - 10\tabcolsep) * \real{0.1667}}
  >{\raggedright\arraybackslash}p{(\linewidth - 10\tabcolsep) * \real{0.1667}}
  >{\raggedright\arraybackslash}p{(\linewidth - 10\tabcolsep) * \real{0.1667}}
  >{\raggedright\arraybackslash}p{(\linewidth - 10\tabcolsep) * \real{0.1667}}@{}}
\caption{Recipient holdout scores, nats/event. † denotes an inadmissible
own fit at the recipient \(b\)-value.}\tabularnewline
\toprule\noalign{}
\begin{minipage}[b]{\linewidth}\raggedright
Recipient
\end{minipage} & \begin{minipage}[b]{\linewidth}\raggedright
Events
\end{minipage} & \begin{minipage}[b]{\linewidth}\raggedright
Clusters
\end{minipage} & \begin{minipage}[b]{\linewidth}\raggedright
Own IG
\end{minipage} & \begin{minipage}[b]{\linewidth}\raggedright
Excluded-pool IG
\end{minipage} & \begin{minipage}[b]{\linewidth}\raggedright
Excluded − own, paired
\end{minipage} \\
\midrule\noalign{}
\endfirsthead
\toprule\noalign{}
\begin{minipage}[b]{\linewidth}\raggedright
Recipient
\end{minipage} & \begin{minipage}[b]{\linewidth}\raggedright
Events
\end{minipage} & \begin{minipage}[b]{\linewidth}\raggedright
Clusters
\end{minipage} & \begin{minipage}[b]{\linewidth}\raggedright
Own IG
\end{minipage} & \begin{minipage}[b]{\linewidth}\raggedright
Excluded-pool IG
\end{minipage} & \begin{minipage}[b]{\linewidth}\raggedright
Excluded − own, paired
\end{minipage} \\
\midrule\noalign{}
\endhead
\bottomrule\noalign{}
\endlastfoot
Mexico & 115 & 91 & +0.698 & +0.720 & +0.022 {[}−0.055, +0.118{]} \\
Central America† & 26 & 21 & +0.531 & +0.678 & +0.146 {[}−0.088,
+0.357{]} \\
Peru--Ecuador† & 99 & 81 & +0.669 & +0.789 & +0.121 {[}−0.058,
+0.291{]} \\
Chile & 147 & 132 & +0.354 & +0.356 & +0.001 {[}−0.021, +0.023{]} \\
Alaska--Aleutians & 57 & 41 & +1.120 & +1.157 & +0.037 {[}−0.138,
+0.225{]} \\
Japan & 182 & 136 & +0.894 & +0.918 & +0.024 {[}−0.007, +0.058{]} \\
Kamchatka--Kuriles & 146 & 76 & +1.855 & +1.896 & +0.042 {[}−0.087,
+0.114{]} \\
Sumatra--Java & 137 & 108 & +1.011 & +1.012 & +0.000 {[}−0.013,
+0.013{]} \\
\end{longtable}

The excluded-pool point estimate is at or above the own fit on eight of
eight recipients, and no paired interval excludes zero. Because \(\mu\)
is re-estimated on the recipient, this design is the one that isolates
the transferred triggering law. The recipients are selected subduction
boxes on the Pacific rim and the Sunda arc with sufficient GCMT training
data, so the evidence supports evaluating pooled laws on further
margins; a universal triggering law, or a guaranteed limit on transfer
loss, would need more.

Two of the own fits are inadmissible, so part of the excluded pool's
showing against them could be the cost of a broken comparator rather
than the value of shared information. To separate the two we score each
recipient also against its own fit refitted inside the admissible
domain, the same restricted fits the likelihood-ratio calibration uses
(\texttt{lr\_admissible.csv}); for the six admissible margins the
restricted fit is the free fit and reproduces its score exactly. The
restriction moves only the two inadmissible comparators, Central America
and Peru--Ecuador, and by at most 0.030 nats/event.

\begin{longtable}[]{@{}
  >{\raggedright\arraybackslash}p{(\linewidth - 8\tabcolsep) * \real{0.2000}}
  >{\raggedright\arraybackslash}p{(\linewidth - 8\tabcolsep) * \real{0.2000}}
  >{\raggedright\arraybackslash}p{(\linewidth - 8\tabcolsep) * \real{0.2000}}
  >{\raggedright\arraybackslash}p{(\linewidth - 8\tabcolsep) * \real{0.2000}}
  >{\raggedright\arraybackslash}p{(\linewidth - 8\tabcolsep) * \real{0.2000}}@{}}
\caption{Leave-one-margin-out against the free own fit and against the
own fit restricted to \(\alpha<\beta\) and \(n<1\). † denotes an
inadmissible free fit; for the others the two comparators
coincide.}\tabularnewline
\toprule\noalign{}
\begin{minipage}[b]{\linewidth}\raggedright
Recipient
\end{minipage} & \begin{minipage}[b]{\linewidth}\raggedright
Own IG
\end{minipage} & \begin{minipage}[b]{\linewidth}\raggedright
Admissible own IG
\end{minipage} & \begin{minipage}[b]{\linewidth}\raggedright
Excluded − own
\end{minipage} & \begin{minipage}[b]{\linewidth}\raggedright
Excluded − admissible
\end{minipage} \\
\midrule\noalign{}
\endfirsthead
\toprule\noalign{}
\begin{minipage}[b]{\linewidth}\raggedright
Recipient
\end{minipage} & \begin{minipage}[b]{\linewidth}\raggedright
Own IG
\end{minipage} & \begin{minipage}[b]{\linewidth}\raggedright
Admissible own IG
\end{minipage} & \begin{minipage}[b]{\linewidth}\raggedright
Excluded − own
\end{minipage} & \begin{minipage}[b]{\linewidth}\raggedright
Excluded − admissible
\end{minipage} \\
\midrule\noalign{}
\endhead
\bottomrule\noalign{}
\endlastfoot
Mexico & +0.698 & +0.698 & +0.022 {[}−0.055, +0.118{]} & +0.022
{[}−0.055, +0.118{]} \\
Central America† & +0.531 & +0.561 & +0.146 {[}−0.088, +0.357{]} &
+0.116 {[}−0.056, +0.269{]} \\
Peru--Ecuador† & +0.669 & +0.669 & +0.121 {[}−0.058, +0.291{]} & +0.120
{[}−0.059, +0.290{]} \\
Chile & +0.354 & +0.354 & +0.001 {[}−0.021, +0.023{]} & +0.001
{[}−0.021, +0.023{]} \\
Alaska--Aleutians & +1.120 & +1.120 & +0.037 {[}−0.138, +0.225{]} &
+0.037 {[}−0.138, +0.225{]} \\
Japan & +0.894 & +0.894 & +0.024 {[}−0.007, +0.058{]} & +0.024
{[}−0.007, +0.058{]} \\
Kamchatka--Kuriles & +1.855 & +1.855 & +0.042 {[}−0.087, +0.114{]} &
+0.042 {[}−0.087, +0.114{]} \\
Sumatra--Java & +1.011 & +1.011 & +0.000 {[}−0.013, +0.013{]} & +0.000
{[}−0.013, +0.013{]} \\
\end{longtable}

Aggregated over the eight recipients with the weighting made explicit,
the event-weighted paired gain of the excluded pool over the admissible
own fits, every holdout event counting once, is +0.0334 {[}+0.0002,
+0.0670{]} nats/event over 909 events in 686 clusters, against +0.0343
{[}+0.0009, +0.0684{]} over the free fits; the unweighted mean of the
eight recipient means is +0.045 {[}+0.017, +0.078{]}, with a bootstrap
over eight recipients that is descriptive. The point estimate is
positive on all eight recipients under either comparator. The
event-weighted lower limit is +0.0002, just above zero, and it is the
interval most exposed to the linking rule: under the nine rules it
excludes zero under seven of nine, with the lower limit running from
−0.0013 to +0.0050 while the point estimate does not move, and the main
rule gives one of the narrowest margins. The coverage of this interval
has not been measured. The 89\% reported above is the coverage of the
SSN gain against Poisson under the fitted SSN law, a different statistic
on a different catalog, and it carries over to this paired contrast
across eight GCMT recipients only as a warning. The aggregate gain is
thus positive with a bootstrap lower limit barely above zero; given the
uncertainty about its coverage we read it neither as conclusive evidence
of superiority nor as a guarantee of noninferiority, and in particular
not as a bound on how much the excluded pool can lose to a well-behaved
local fit. The restriction does show that the aggregate barely depends
on the two inadmissible comparators.

\subsection{Partial pooling}\label{partial-pooling}

We shrink recipient parameters toward the law fitted without that
recipient, using a Gaussian penalty on \(\log K\), \(\log\alpha\),
\(\log c\), \(\log(p-1)\), \(\log d\) and \(\log(q-1)\). \(\mu\) is
unpenalized. The width is \(\lambda\) times the between-margin standard
deviation among admissible donor fits, excluding the recipient.
\(\lambda=0\) is fully pooled and \(\lambda\) tending to infinity is the
free fit. The empirical dispersion is an ad hoc scale that includes
estimation noise; a hierarchical model would estimate the population
variance jointly.

\begin{longtable}[]{@{}llllll@{}}
\caption{Partial pooling, event-weighted across 909 recipient holdout
events. ``Above own'' counts recipients beating their own fit by more
than \(10^{-6}\) nats/event. Per-recipient paired intervals are in the
supplement.}\tabularnewline
\toprule\noalign{}
\(\lambda\) & IG & vs own & vs pooled & Above own & Inadmissible \\
\midrule\noalign{}
\endfirsthead
\toprule\noalign{}
\(\lambda\) & IG & vs own & vs pooled & Above own & Inadmissible \\
\midrule\noalign{}
\endhead
\bottomrule\noalign{}
\endlastfoot
0 & +0.967 & +0.034 & +0.000 & 8 & 0 \\
0.25 & +0.962 & +0.029 & −0.005 & 6 & 0 \\
0.5 & +0.959 & +0.025 & −0.009 & 6 & 0 \\
1 & +0.953 & +0.020 & −0.015 & 5 & 0 \\
2 & +0.944 & +0.011 & −0.023 & 5 & 1 \\
4 & +0.939 & +0.006 & −0.029 & 5 & 1 \\
$\infty$ & +0.933 & +0.000 & −0.034 & --- & 2 \\
\end{longtable}

The infinite-width row is the free fit by construction, so it has no
meaningful count against itself; every recipient there reproduces its
own score to within \(10^{-5}\) nats/event. Counts at finite widths
should be read with their magnitudes: at width zero the eight recipients
include one whose margin over its own fit is 0.0005 nats/event, far
below any resolvable difference.

The entire width sweep is reported; no width is chosen by selecting the
best holdout result. Event weighting differs from equal weighting of
margins, particularly because the two inadmissible recipients have
relatively small holdouts. Conclusions about the middle of the family
rest on these paired scores, since parameter shrinkage alone does not
measure forecast value.

\section{Rolling origins: annual performance and estimator
variation}\label{rolling-origins-annual-performance-and-estimator-variation}

At origins spaced every six months from January 2012 through July 2025,
training uses {[}2000,T), with the first year as conditioning history,
and scoring uses {[}T,T+365.25 days). The replay runs on the final
catalog; no forecasts were archived in real time. The one-year windows
overlap, so their totals must not be summed and called performance on a
unique fourteen-year history. Background maps and rates are re-estimated
within each origin.

A refits all seven local SSN parameters at magnitude 5.0. B transfers
the pooled law to SSN magnitude 5.5 and therefore scores a different
target. C freezes the pooled triggering law and re-estimates only local
\(\mu\) at magnitude 5.0. D freezes the local law fitted before the
first origin and re-estimates \(\mu\). E refits locally with the
shrinkage penalty centered on C, at prior width \(\lambda=1\). Width 1
is the unit of the penalty's own scale, one between-margin standard
deviation, and was fixed in advance as the natural midpoint of the
family, before the rolling results existed; the full width sweep appears
in the partial-pooling section on the GCMT recipients. Every pooled law
in this comparison is fitted on GCMT data ending in 2012; the full
transfer analysis uses the 2016 law. E's prior widths and donor
admissibility selection also use only pre-2012 fits, excluding Mexico
from the dispersion estimate.

B is scored at 27 of the 28 origins. At the first origin its training
window holds 56 events above SSN magnitude 5.5, below the 60-event
selection floor used for the GCMT margins, so that fold is not run. The
floor is a data-sufficiency rule; whether seven parameters are
identified above it is what the identifiability section reports. The
floor is applied identically to every variant; only B, whose higher
threshold roughly quarters the event count, is affected by it.

Two calibration choices remain retrospective, and they are retrospective
in different senses. The SSN--Mw offset of 0.09 was estimated from
matched events over the full available record rather than reconstructed
at each origin, and the offset sweep below measures sensitivity to one
constant calibration error. The Gutenberg--Richter slope of 0.95 was not
estimated from this record at all; it is a fixed convention, and the
following section reports what the measured alternatives do to these
results. Neither sweep makes the replay fully prospective or tests a
magnitude-dependent or time-varying conversion.

\begin{longtable}[]{@{}
  >{\raggedright\arraybackslash}p{(\linewidth - 14\tabcolsep) * \real{0.1250}}
  >{\raggedright\arraybackslash}p{(\linewidth - 14\tabcolsep) * \real{0.1250}}
  >{\raggedright\arraybackslash}p{(\linewidth - 14\tabcolsep) * \real{0.1250}}
  >{\raggedright\arraybackslash}p{(\linewidth - 14\tabcolsep) * \real{0.1250}}
  >{\raggedright\arraybackslash}p{(\linewidth - 14\tabcolsep) * \real{0.1250}}
  >{\raggedright\arraybackslash}p{(\linewidth - 14\tabcolsep) * \real{0.1250}}
  >{\raggedright\arraybackslash}p{(\linewidth - 14\tabcolsep) * \real{0.1250}}
  >{\raggedright\arraybackslash}p{(\linewidth - 14\tabcolsep) * \real{0.1250}}@{}}
\caption{Rolling summary. Gains are nats; windows won means positive
total gain over Poisson. B scores a different magnitude target and is
not ranked against the other variants.}\tabularnewline
\toprule\noalign{}
\begin{minipage}[b]{\linewidth}\raggedright
Variant
\end{minipage} & \begin{minipage}[b]{\linewidth}\raggedright
Origins
\end{minipage} & \begin{minipage}[b]{\linewidth}\raggedright
Events / year
\end{minipage} & \begin{minipage}[b]{\linewidth}\raggedright
Mean / event
\end{minipage} & \begin{minipage}[b]{\linewidth}\raggedright
Median / event
\end{minipage} & \begin{minipage}[b]{\linewidth}\raggedright
Mean / year
\end{minipage} & \begin{minipage}[b]{\linewidth}\raggedright
Windows won
\end{minipage} & \begin{minipage}[b]{\linewidth}\raggedright
\(K\) range
\end{minipage} \\
\midrule\noalign{}
\endfirsthead
\toprule\noalign{}
\begin{minipage}[b]{\linewidth}\raggedright
Variant
\end{minipage} & \begin{minipage}[b]{\linewidth}\raggedright
Origins
\end{minipage} & \begin{minipage}[b]{\linewidth}\raggedright
Events / year
\end{minipage} & \begin{minipage}[b]{\linewidth}\raggedright
Mean / event
\end{minipage} & \begin{minipage}[b]{\linewidth}\raggedright
Median / event
\end{minipage} & \begin{minipage}[b]{\linewidth}\raggedright
Mean / year
\end{minipage} & \begin{minipage}[b]{\linewidth}\raggedright
Windows won
\end{minipage} & \begin{minipage}[b]{\linewidth}\raggedright
\(K\) range
\end{minipage} \\
\midrule\noalign{}
\endhead
\bottomrule\noalign{}
\endlastfoot
A: local refitted & 28 & 30.1 & +0.716 & +0.733 & +29.55 & 79\% &
5.40 \\
B: pooled, SSN 5.5 & 27 & 8.0 & +0.502 & +0.312 & +5.65 & 67\% & 1.00 \\
C: pooled frozen & 28 & 30.1 & +0.728 & +0.602 & +31.61 & 79\% & 1.00 \\
D: local frozen & 28 & 30.1 & +0.697 & +0.729 & +28.01 & 79\% & 1.00 \\
E: local shrunk & 28 & 30.1 & +0.718 & +0.711 & +29.77 & 79\% & 3.37 \\
\end{longtable}

\begin{longtable}[]{@{}
  >{\raggedright\arraybackslash}p{(\linewidth - 8\tabcolsep) * \real{0.2000}}
  >{\raggedright\arraybackslash}p{(\linewidth - 8\tabcolsep) * \real{0.2000}}
  >{\raggedright\arraybackslash}p{(\linewidth - 8\tabcolsep) * \real{0.2000}}
  >{\raggedright\arraybackslash}p{(\linewidth - 8\tabcolsep) * \real{0.2000}}
  >{\raggedright\arraybackslash}p{(\linewidth - 8\tabcolsep) * \real{0.2000}}@{}}
\caption{Paired moving-block intervals at block length 4, on differences
formed origin by origin. Annual gain is the primary operational
estimand; all lengths 1--8 are supplied.}\tabularnewline
\toprule\noalign{}
\begin{minipage}[b]{\linewidth}\raggedright
Contrast
\end{minipage} & \begin{minipage}[b]{\linewidth}\raggedright
Mean annual gain
\end{minipage} & \begin{minipage}[b]{\linewidth}\raggedright
Median annual gain
\end{minipage} & \begin{minipage}[b]{\linewidth}\raggedright
Mean per-event gain
\end{minipage} & \begin{minipage}[b]{\linewidth}\raggedright
Median per-event gain
\end{minipage} \\
\midrule\noalign{}
\endfirsthead
\toprule\noalign{}
\begin{minipage}[b]{\linewidth}\raggedright
Contrast
\end{minipage} & \begin{minipage}[b]{\linewidth}\raggedright
Mean annual gain
\end{minipage} & \begin{minipage}[b]{\linewidth}\raggedright
Median annual gain
\end{minipage} & \begin{minipage}[b]{\linewidth}\raggedright
Mean per-event gain
\end{minipage} & \begin{minipage}[b]{\linewidth}\raggedright
Median per-event gain
\end{minipage} \\
\midrule\noalign{}
\endhead
\bottomrule\noalign{}
\endlastfoot
A - C & −2.07 {[}−6.92, +1.15{]} & +0.73 {[}−0.37, +1.56{]} & −0.011
{[}−0.083, +0.046{]} & +0.038 {[}−0.017, +0.057{]} \\
A - D & +1.54 {[}−0.16, +4.55{]} & −0.09 {[}−0.35, +1.35{]} & +0.019
{[}−0.011, +0.067{]} & −0.003 {[}−0.019, +0.050{]} \\
D - C & −3.60 {[}−11.28, +1.11{]} & +0.80 {[}−0.98, +1.13{]} & −0.030
{[}−0.148, +0.052{]} & +0.029 {[}−0.031, +0.058{]} \\
E - A & +0.23 {[}−0.19, +0.72{]} & −0.01 {[}−0.22, +0.25{]} & +0.001
{[}−0.009, +0.010{]} & −0.001 {[}−0.012, +0.009{]} \\
E - C & −1.84 {[}−6.21, +1.00{]} & +0.49 {[}−0.23, +1.20{]} & −0.010
{[}−0.075, +0.039{]} & +0.028 {[}−0.009, +0.040{]} \\
E - D & +1.76 {[}−0.31, +5.22{]} & −0.12 {[}−0.56, +1.77{]} & +0.021
{[}−0.017, +0.075{]} & −0.005 {[}−0.030, +0.055{]} \\
\end{longtable}

These are paired contrasts. For the means the pairing changes nothing:
the same 28 origins enter with equal weight, so the paired mean
per-event A−C gain of −0.011 equals the difference of the two
candidates' means, −0.011. For the medians it does: the paired median
per-event A−C gain is +0.038, whereas the two candidates' own medians
differ by +0.131, and the mean annual and median annual A−C contrasts
carry opposite signs. Reporting a paired median alone, or subtracting
one table's medians from the other's, would misstate the comparison.

The block intervals do not reach their nominal level. Under the frozen
pooled law the length-4 interval on C's own mean annual gain covers in
87\% of 200 synthetic replays, and on the paired contrast D − C in 76\%,
with the median interval at 80\%. Length 4 is close to the best
available: length 1, the ordinary bootstrap, covers the paired mean in
70\%, and longer blocks lose coverage again as the number of blocks
falls. The shortfall on the paired contrast is structural, and no block
length fixes it: its interval half-width averages 0.50 nats/year while
the contrast's standard deviation across replays is 0.46, because the
estimation error of a law fitted once per catalog is common to all 28
origins and resampling origins within a catalog cannot see it. This
measurement belongs to D−C. A−C and E−C share the frozen comparator but
not the estimator on the other side: D is estimated once per catalog, A
is refitted at every origin and E adds a penalty, and sharing evaluation
windows does not make their error distributions the same. Their coverage
has not been measured. The D−C experiment warns of undercoverage in the
primary comparisons, whose intervals rest on the same resampling of
origins, without quantifying it. Measuring undercoverage would not
correct the intervals either: calibrating them requires an additional
procedure, such as bootstrap calibration of the nominal level, and a
check of that procedure's own performance (\citeproc{ref-Loh1991}{Loh
1991}), neither of which we have done. The lag-1 autocorrelation of the
simulated D − C series, +0.46, is close to the observed +0.52, which
confirms that the dependence the blocks address is real; what they miss
is the shared component.

\begin{longtable}[]{@{}llllll@{}}
\caption{Coverage of the 95\% moving-block intervals under the frozen
pooled law, 200 synthetic catalogs replayed through the 28 origins.
Columns are the mean interval at block lengths 1, 2, 4 and 8, and the
median interval at length 4.}\tabularnewline
\toprule\noalign{}
Contrast & L = 1 & L = 2 & L = 4 & L = 8 & median, L = 4 \\
\midrule\noalign{}
\endfirsthead
\toprule\noalign{}
Contrast & L = 1 & L = 2 & L = 4 & L = 8 & median, L = 4 \\
\midrule\noalign{}
\endhead
\bottomrule\noalign{}
\endlastfoot
C, annual & 83\% & 86\% & 87\% & 84\% & 88\% \\
C, per event & 83\% & 88\% & 88\% & 86\% & 88\% \\
D - C, annual & 70\% & 78\% & 76\% & 70\% & 80\% \\
D - C, per event & 72\% & 76\% & 77\% & 72\% & 79\% \\
\end{longtable}

Mean and median intervals need not have the same relation to zero. A
candidate can outperform another in a typical quiet year and lose more
during a small number of high-count sequence years. A resolved median
contrast is reported as such and says nothing about mean annual utility;
conversely, an unresolved mean does not show that the candidates have
the same skill. The choice of statistic should precede the observed
ranking.

Freezing guarantees constant triggering coefficients. It does not
guarantee unchanged forecasts: new conditioning earthquakes,
re-estimated \(\mu\) and revised spatial backgrounds change the
intensity field. A direct comparison of forecast discontinuities would
hold the event history and issue time fixed while swapping old and new
complete model versions. We have not measured that cost, and
parameter-coordinate dispersion is only an indirect diagnostic.

A's productivity ranges by a factor of 5.40, with a maximum
adjacent-origin ratio of 2.73. E's range is 3.37. These ranges describe
the estimator's trajectory; they are not confidence intervals, and they
say nothing about whether the physical law changed. Both frozen laws
have zero triggering-parameter dispersion by construction, so the
stability is a property of freezing and applies equally to the local and
the pooled law.

\begin{figure}
\centering
\includegraphics[width=1\linewidth,height=\textheight,keepaspectratio,alt={Rolling gains and productivity. A and E update triggering parameters; C and D keep them fixed. Panel (b) uses a logarithmic K axis. The figure does not measure discontinuities caused by background updates. Paired annual, mean-per-event and median-per-event contrasts are reported separately.}]{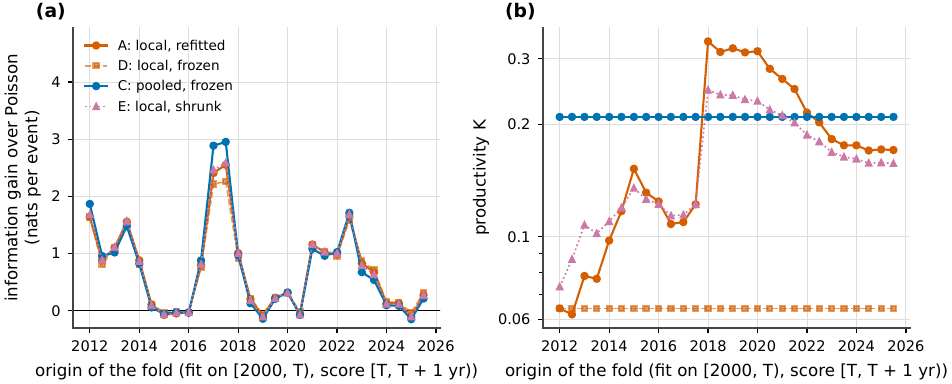}
\caption{Rolling gains and productivity. A and E update triggering
parameters; C and D keep them fixed. Panel (b) uses a logarithmic \(K\)
axis. The figure does not measure discontinuities caused by background
updates. Paired annual, mean-per-event and median-per-event contrasts
are reported separately.}
\end{figure}

\subsection{Matched freezing dates}\label{matched-freezing-dates}

We compare fits ending in 2012 and 2016, re-reference both pooled laws
to the same SSN threshold 5.0, and evaluate the same SSN-magnitude
parent. For an M6 parent over 3,650 days and 400 km, ignoring
domain-edge losses, the local law's predicted direct offspring change by
+89.75\% and the pooled law's by -0.31\%. Both percentages refer to the
same parent and offspring thresholds. The complete magnitude/horizon
grid is supplied; percentages from native Mw 6.5 and SSN M6.0
calculations are not interchangeable.

\begin{longtable}[]{@{}lll@{}}
\caption{2012-to-2016 coordinate changes, both laws expressed at the SSN
threshold.}\tabularnewline
\toprule\noalign{}
Parameter & Pooled change & Local change \\
\midrule\noalign{}
\endfirsthead
\toprule\noalign{}
Parameter & Pooled change & Local change \\
\midrule\noalign{}
\endhead
\bottomrule\noalign{}
\endlastfoot
\(K\) & +3.87\% & +93.52\% \\
\(\alpha\) & -2.54\% & +26.23\% \\
\(c\) & -19.01\% & -62.60\% \\
\(p\) & -0.81\% & -8.97\% \\
\(d\) & -9.47\% & +1.83\% \\
\(q\) & -1.48\% & -10.00\% \\
\end{longtable}

This is a single matched interval; a freezing-date sensitivity curve
would need other endpoints, which could be estimated. The pooled windows
share 32 of 36 scored years, whereas the local windows share 11 of 15,
so training-history length contributes to the difference in stability.
Coordinate reductions differ by parameter, as the table shows.

\section{Magnitude-threshold conversion and synthetic
checks}\label{magnitude-threshold-conversion-and-synthetic-checks}

The pooled law is fitted at Mw 5.5. We approximate SSN magnitude as
Mw−0.09 using a constant offset derived from matched events. Thus an SSN
threshold Mc corresponds to Mw \(M^*=M_c+0.09\). Under an unbounded
Gutenberg--Richter scaling assumption,

\[K_{\rm eff}=K_G\exp[(\beta-\alpha)(5.5-M_c-0.09)],\qquad\beta=b\ln10.\]

The offset cancels inside the parent-minus-threshold magnitude
difference; the threshold transformation retains \(\beta-\alpha\), not
\(\alpha\) alone. The slope \(\beta\) in this expression is a property
of the \textbf{recipient} catalog, because the factor counts offspring
above the SSN threshold per offspring above Mw 5.5: it is the
Gutenberg--Richter slope of the SSN population between the two
thresholds, not the donor GCMT slope and not a free constant. This
algebra does not make an observed high-threshold ETAS process invariant
to admitting hidden lower-magnitude parents. Unobserved cascades can
alter apparent productivity, background and kernel shape
(\citeproc{ref-Sornette2005}{Sornette and Werner 2005};
\citeproc{ref-Saichev2006}{Saichev and Sornette 2006}). With an upper
magnitude bound, even the direct threshold-survival ratio requires its
truncation normalization for literal algebraic exactness; we retain the
unbounded formula as the operational approximation tested here.

At \(b=0.95\) the 2016 law gives \(K_{\rm eff}=0.2175\), reference
\(n=0.516\) and cutoff-only \(n_T=0.441\). The independently fitted
masked SSN law has \(K=0.2151\). Agreement or disagreement between these
estimates is not a validation of the conversion, because their training
data, observation processes and inferred kernels differ.

The offset sensitivity sweep uses the \textbf{2012} law throughout, with
offsets from -0.19 to 0.01. Its \(K_{\rm eff}\) range is 0.1916--0.2289,
and mean annual per-event gains range from 0.7228 to 0.7303. It tests
constant-offset sensitivity only. Changing a magnitude-conversion slope
would change relative parent productivities and is outside this sweep.

\subsection{The assumed Gutenberg--Richter
slope}\label{the-assumed-gutenbergrichter-slope}

The value \(b=0.95\) used throughout this paper is a fixed convention.
The measured values at the adopted threshold were given in the data
section, 1.020 for the whole rectangle and 1.106 for the on-slab
training subset, and the convention sits 2.0 standard errors below the
first and 3.3 below the second. The nearest quantity anywhere in the
diagnostic tables is the least-squares cross-check slope of 0.949 at the
magnitude-4.5 cutoff, and that cutoff is the one this paper rejects as
misspecified.

The convention is retained because \(b\) enters several reported
quantities at once, and this section reports its influence. It sets
\(K_{\rm eff}\) through the threshold conversion, so it moves the pooled
law's productivity at the SSN threshold; it appears in the reference
branching ratio, where it very largely cancels; and it is the slope from
which simulated magnitudes are drawn in the count and synthetic
experiments. The likelihood-ratio bootstrap does not use it: each margin
is simulated with its own maximum-likelihood slope from the free fits,
0.78 to 0.87, because \(b\) is not a triggering parameter and a
per-margin null should reproduce each catalog's own magnitude
distribution.

The following sweep substitutes each measured value for the convention,
recomputes the re-referenced 2012 law, and reruns both variants that
depend on it, C and E, over all 28 rolling origins. Variants A and D
never see \(b\), so their folds are reused unchanged and every contrast
stays paired on the same origins. E is worth separating from C: \(b\)
enters it through a shrinkage penalty rather than through the law being
scored, and it moves the prior center while leaving the prior width
alone, since that width is the between-margin dispersion of the GCMT
free fits and knows nothing about the SSN magnitude distribution. A
center that moves against a fixed width changes how hard the prior
pulls, not only where it pulls, so the effect on E need not follow the
effect on C. At the convention the sweep reproduces the variant E folds
of the partial-pooling run to within \(7.6\times10^{-17}\) nats per
event at the worst origin.

\begin{longtable}[]{@{}
  >{\raggedright\arraybackslash}p{(\linewidth - 10\tabcolsep) * \real{0.1667}}
  >{\raggedright\arraybackslash}p{(\linewidth - 10\tabcolsep) * \real{0.1667}}
  >{\raggedright\arraybackslash}p{(\linewidth - 10\tabcolsep) * \real{0.1667}}
  >{\raggedright\arraybackslash}p{(\linewidth - 10\tabcolsep) * \real{0.1667}}
  >{\raggedright\arraybackslash}p{(\linewidth - 10\tabcolsep) * \real{0.1667}}
  >{\raggedright\arraybackslash}p{(\linewidth - 10\tabcolsep) * \real{0.1667}}@{}}
\caption{Sensitivity of the re-referenced 2012 pooled law and the
primary rolling contrast to the assumed Gutenberg--Richter slope.
Intervals are paired moving-block at length 4.}\tabularnewline
\toprule\noalign{}
\begin{minipage}[b]{\linewidth}\raggedright
\(b\) (source)
\end{minipage} & \begin{minipage}[b]{\linewidth}\raggedright
\(K_{\rm eff}\)
\end{minipage} & \begin{minipage}[b]{\linewidth}\raggedright
\(n\)
\end{minipage} & \begin{minipage}[b]{\linewidth}\raggedright
\(n_T\)
\end{minipage} & \begin{minipage}[b]{\linewidth}\raggedright
A−C mean annual
\end{minipage} & \begin{minipage}[b]{\linewidth}\raggedright
A−C median per event
\end{minipage} \\
\midrule\noalign{}
\endfirsthead
\toprule\noalign{}
\begin{minipage}[b]{\linewidth}\raggedright
\(b\) (source)
\end{minipage} & \begin{minipage}[b]{\linewidth}\raggedright
\(K_{\rm eff}\)
\end{minipage} & \begin{minipage}[b]{\linewidth}\raggedright
\(n\)
\end{minipage} & \begin{minipage}[b]{\linewidth}\raggedright
\(n_T\)
\end{minipage} & \begin{minipage}[b]{\linewidth}\raggedright
A−C mean annual
\end{minipage} & \begin{minipage}[b]{\linewidth}\raggedright
A−C median per event
\end{minipage} \\
\midrule\noalign{}
\endhead
\bottomrule\noalign{}
\endlastfoot
0.950 (fixed convention) & 0.2094 & 0.515 & 0.444 & −2.07 {[}−6.92,
+1.15{]} & +0.038 {[}−0.017, +0.057{]} \\
1.020 (whole rectangle, n=961) & 0.2237 & 0.500 & 0.431 & −2.09
{[}−7.19, +1.27{]} & +0.040 {[}−0.005, +0.060{]} \\
1.106 (on-slab subset, n=545) & 0.2426 & 0.495 & 0.426 & −2.08 {[}−7.46,
+1.48{]} & +0.045 {[}+0.009, +0.065{]} \\
\end{longtable}

\begin{longtable}[]{@{}
  >{\raggedright\arraybackslash}p{(\linewidth - 10\tabcolsep) * \real{0.1667}}
  >{\raggedright\arraybackslash}p{(\linewidth - 10\tabcolsep) * \real{0.1667}}
  >{\raggedright\arraybackslash}p{(\linewidth - 10\tabcolsep) * \real{0.1667}}
  >{\raggedright\arraybackslash}p{(\linewidth - 10\tabcolsep) * \real{0.1667}}
  >{\raggedright\arraybackslash}p{(\linewidth - 10\tabcolsep) * \real{0.1667}}
  >{\raggedright\arraybackslash}p{(\linewidth - 10\tabcolsep) * \real{0.1667}}@{}}
\caption{Variant E under the same sweep. Its prior center is the
re-referenced law and moves with \(b\); its prior width does
not.}\tabularnewline
\toprule\noalign{}
\begin{minipage}[b]{\linewidth}\raggedright
\(b\)
\end{minipage} & \begin{minipage}[b]{\linewidth}\raggedright
E mean / event
\end{minipage} & \begin{minipage}[b]{\linewidth}\raggedright
E \(K\) range
\end{minipage} & \begin{minipage}[b]{\linewidth}\raggedright
E − A mean annual
\end{minipage} & \begin{minipage}[b]{\linewidth}\raggedright
E − C mean annual
\end{minipage} & \begin{minipage}[b]{\linewidth}\raggedright
E − C median per event
\end{minipage} \\
\midrule\noalign{}
\endfirsthead
\toprule\noalign{}
\begin{minipage}[b]{\linewidth}\raggedright
\(b\)
\end{minipage} & \begin{minipage}[b]{\linewidth}\raggedright
E mean / event
\end{minipage} & \begin{minipage}[b]{\linewidth}\raggedright
E \(K\) range
\end{minipage} & \begin{minipage}[b]{\linewidth}\raggedright
E − A mean annual
\end{minipage} & \begin{minipage}[b]{\linewidth}\raggedright
E − C mean annual
\end{minipage} & \begin{minipage}[b]{\linewidth}\raggedright
E − C median per event
\end{minipage} \\
\midrule\noalign{}
\endhead
\bottomrule\noalign{}
\endlastfoot
0.950 & +0.718 & 3.37 & +0.23 {[}−0.19, +0.72{]} & −1.84 {[}−6.21,
+1.00{]} & +0.028 {[}−0.009, +0.040{]} \\
1.020 & +0.718 & 3.35 & +0.25 {[}−0.18, +0.76{]} & −1.85 {[}−6.45,
+1.13{]} & +0.032 {[}+0.000, +0.043{]} \\
1.106 & +0.718 & 3.33 & +0.27 {[}−0.18, +0.80{]} & −1.81 {[}−6.67,
+1.34{]} & +0.037 {[}+0.018, +0.051{]} \\
\end{longtable}

\(K_{\rm eff}\) rises by 16\% between the convention and the on-slab
measurement, because a steeper magnitude distribution implies more
offspring below the lower threshold. The reference branching ratio
barely responds, moving from 0.515 to 0.495, since the same \(\beta\)
multiplies the numerator and divides the denominator; the admissibility
guard and the ratios reported against it are therefore effectively free
of this choice. Variant C's own performance is almost unmoved in the
aggregate: it wins 79\% of years at every value and its mean annual gain
ranges from +31.613 to +31.640 nats per year across the three slopes, a
spread of 0.028. Its mean per-event gain does drift down slightly, which
matters for the contrasts and is taken up below.

The primary contrast is stable. A−C in mean annual gain is −2.07 at the
convention and −2.08 at the on-slab value, with intervals including zero
throughout, so the finding that local refitting does not resolve an
annual advantage does not depend on \(b\).

One reported result does move: the paired \textbf{median} contrast
sharpens as \(b\) rises. Per event it goes from +0.038 {[}−0.017,
+0.057{]} at the convention to +0.045 {[}+0.009, +0.065{]} at the
on-slab value, and the annual median from +0.73 {[}−0.37, +1.56{]} to
+1.16 {[}+0.22, +2.00{]}. The per-event interval includes zero at the
convention and the whole-rectangle slope (1.020) and excludes it at the
on-slab slope (1.106). The statement that the paired median is positive
by an unresolved margin is thus conditional on \(b\), whereas the
statement that the mean annual contrast is unresolved holds across the
range. We do not treat the resolved version as establishing an
operational advantage. The coverage of this interval is unmeasured, and
the one paired block interval whose coverage was measured, D−C under the
pooled null, fell well short of its nominal level; three slopes were
examined and this is the result at one of them; and no noninferiority
margin was predefined. The median evidence is stronger under the
measured slope than under the convention, and a deployment argument
resting on the median would have to be made at a defensible \(b\).

Variant E behaves the same way, and its own score barely notices the
change. Its mean per-event gain moves from +0.7177 to +0.7184 and its
median from +0.7114 to +0.7113, changes small relative to the reported
interval widths. Its productivity dispersion is equally steady, a range
of 3.37 at the convention against 3.33 at the on-slab value, so the
finding that shrinkage roughly halves the refitted estimator's
dispersion without removing it does not rest on the assumed slope. The
one E coordinate that does respond is its largest reference branching
ratio, which falls from 0.587 to 0.496, following the prior center it is
shrunk toward. E − A is unresolved at every value, from +0.23 {[}−0.19,
+0.72{]} to +0.27 {[}−0.18, +0.80{]} nats per year.

E's insensitivity locates the mechanism. The prior center moves with
\(b\) while the prior width does not, so the penalty pulls toward a
different point with unchanged strength; E absorbs that almost entirely,
because \(\mu\) is re-estimated per fold and the six penalized
coordinates are free to follow the data. What moves in the E − C
contrast is C, and the same holds for A − C, where A cannot move at all.
Across the sweep C's mean per-event gain declines from +0.7276 to
+0.7186 while both refitting variants stay put, and that decline is the
whole of the widening in both contrasts.

The consequence is that the median result is not a peculiarity of one
comparison. E − C moves with A − C, though not identically: +0.028
{[}−0.009, +0.040{]} per event at the convention, +0.037 {[}+0.018,
+0.051{]} at the on-slab value, and annually +0.49 {[}−0.23, +1.20{]}
against +0.92 {[}+0.36, +1.59{]}. The E − C per-event median excludes
zero at the whole-rectangle slope (1.020) and the on-slab slope (1.106),
at the whole-rectangle slope by a lower limit of +0.0002, and includes
it at the convention; A − C excludes it at the on-slab slope (1.106)
only. Where the two differ, at the whole-rectangle slope, we do not read
the E − C interval as resolving the contrast, and the reasons are these.
Its coverage is unmeasured, and the only paired block interval whose
coverage was measured fell short of nominal. Three slopes were examined
for two contrasts, and this is the narrowest margin among the six
intervals, of which three exclude zero. And the interval does not
establish a practically meaningful minimum improvement. Both mean annual
contrasts have negative point estimates and intervals including zero at
every value. Two differently constructed locally adapted laws, one
freely refitted and one shrunk toward the pooled law, behave alike:
their paired medians against the frozen pooled law are positive at every
\(b\), A − C excludes zero at the on-slab slope (1.106) and E − C at the
whole-rectangle slope (1.020) and the on-slab slope (1.106), and their
mean annual contrasts are negative and unresolved at every \(b\). We
report both.

The simulated-magnitude uses of \(b\) are not swept. Raising \(b\) there
would lower the mean simulated magnitude and slightly reduce simulated
cascade sizes, which bears on the count and likelihood-ratio experiments
but not on the observed-data scores that carry the paper's conclusions.

The synthetic study uses the pooled kernel shapes, \(b=0.95\), a uniform
rectangular background, five years of conditioning history, thirty years
of training and ten years of holdout. Five cases have ten seeds each.
All use the fitted interaction cutoffs and continuous magnitude
truncation at 8.4. The control permits only parents at Mw 5.5 and above
to trigger; its low-threshold catalog is not ordinary ETAS. The
remaining cases include hidden-parent cascades below one or both fitting
thresholds. Fits and the re-referenced candidate are evaluated on the
same low-threshold holdout, with \(\mu\) re-estimated. These simulations
use a simpler background than the real masked KDE and a uniform
completeness.

\begin{longtable}[]{@{}
  >{\raggedright\arraybackslash}p{(\linewidth - 8\tabcolsep) * \real{0.2000}}
  >{\raggedright\arraybackslash}p{(\linewidth - 8\tabcolsep) * \real{0.2000}}
  >{\raggedright\arraybackslash}p{(\linewidth - 8\tabcolsep) * \real{0.2000}}
  >{\raggedright\arraybackslash}p{(\linewidth - 8\tabcolsep) * \real{0.2000}}
  >{\raggedright\arraybackslash}p{(\linewidth - 8\tabcolsep) * \real{0.2000}}@{}}
\caption{Synthetic threshold check, all seeds including flagged fits.
Ratios compare re-referenced \(K\) with the algebraic control value or
direct low-threshold fit. Brackets are seed ranges; they are not
confidence intervals. Unresolved counts include nonconvergence or
failure of the stated interior/reference guard.}\tabularnewline
\toprule\noalign{}
\begin{minipage}[b]{\linewidth}\raggedright
Generating case
\end{minipage} & \begin{minipage}[b]{\linewidth}\raggedright
\(K\) ratio, median {[}range{]}
\end{minipage} & \begin{minipage}[b]{\linewidth}\raggedright
Paired gain, median {[}range{]}
\end{minipage} & \begin{minipage}[b]{\linewidth}\raggedright
High-fit fallback
\end{minipage} & \begin{minipage}[b]{\linewidth}\raggedright
Unresolved hi / lo
\end{minipage} \\
\midrule\noalign{}
\endfirsthead
\toprule\noalign{}
\begin{minipage}[b]{\linewidth}\raggedright
Generating case
\end{minipage} & \begin{minipage}[b]{\linewidth}\raggedright
\(K\) ratio, median {[}range{]}
\end{minipage} & \begin{minipage}[b]{\linewidth}\raggedright
Paired gain, median {[}range{]}
\end{minipage} & \begin{minipage}[b]{\linewidth}\raggedright
High-fit fallback
\end{minipage} & \begin{minipage}[b]{\linewidth}\raggedright
Unresolved hi / lo
\end{minipage} \\
\midrule\noalign{}
\endhead
\bottomrule\noalign{}
\endlastfoot
control: only M\textgreater=5.5 triggers & 0.985 {[}0.575, 1.460{]} &
not an ETAS comparison & 1/10 & 0 / 0 \\
ETAS from 5.09 & 1.198 {[}0.966, 1.557{]} & −0.002 {[}−0.013, +0.000{]}
& 0/10 & 0 / 0 \\
ETAS from 4.5, \(n=0.7\) & 1.356 {[}0.993, 3.475{]} & −0.003 {[}−0.011,
+0.001{]} & 1/10 & 1 / 0 \\
ETAS from 4.5, \(n=0.85\) & 1.426 {[}1.096, 2.136{]} & −0.014 {[}−0.039,
+0.007{]} & 2/10 & 2 / 1 \\
ETAS from 4.0, \(n=0.7\) & 1.608 {[}0.647, \(2.12\times10^{14}\){]} &
−0.004 {[}−0.017, +0.008{]} & 4/10 & 4 / 1 \\
\end{longtable}

Some simulated fits reach a near-boundary Omori ridge or fail the
reference guard. The extreme upper \(K\) ratio in the Mw 4.0 case comes
from an unresolved boundary fit and is not interpretable as recovered
physical productivity. The recovery procedure attempts direct
maximization from additional starts and selects an interior admissible
solution when one is found; fallback flags and likelihood differences
are reported. This selection is part of the estimator being tested and
should not be confused with unconditional recovery by a globally
maximized unconstrained likelihood. The seed ranges measure behavior
under the selected generating cases; they say nothing about forecast
error in general or about the cause of any particular real-data
discrepancy between pooled and local fits.

\section{Local corrections to the pooled
law}\label{local-corrections-to-the-pooled-law}

On the SSN split, we free specified subsets of triggering parameters
while re-estimating \(\mu\) in every candidate. All are compared with
the pinned, re-referenced 2016 pooled law on the same 101 holdout
events. No attribution was prespecified, so the ladder is exploratory.
Kernel and productivity parameters interact: a change in \(c\) or \(p\)
can change retained productivity without changing \(K\). No causal
attribution follows merely from whichever subset has the largest point
estimate.

\begin{longtable}[]{@{}
  >{\raggedright\arraybackslash}p{(\linewidth - 14\tabcolsep) * \real{0.1250}}
  >{\raggedright\arraybackslash}p{(\linewidth - 14\tabcolsep) * \real{0.1250}}
  >{\raggedright\arraybackslash}p{(\linewidth - 14\tabcolsep) * \real{0.1250}}
  >{\raggedright\arraybackslash}p{(\linewidth - 14\tabcolsep) * \real{0.1250}}
  >{\raggedright\arraybackslash}p{(\linewidth - 14\tabcolsep) * \real{0.1250}}
  >{\raggedright\arraybackslash}p{(\linewidth - 14\tabcolsep) * \real{0.1250}}
  >{\raggedright\arraybackslash}p{(\linewidth - 14\tabcolsep) * \real{0.1250}}
  >{\raggedright\arraybackslash}p{(\linewidth - 14\tabcolsep) * \real{0.1250}}@{}}
\caption{Local correction ladder, nats/event. Branching ratios use the
fixed \(b=0.95\). For the re-referenced pooled law the ratio is
insensitive to \(b\) (0.515 to 0.495 across the measured alternatives)
because \(K_{\rm eff}\) is transformed with it; for these local rungs
\(K\) is not, and the reference ratio of the full local fit moves from
0.565 at \(b=0.95\) to 0.460 at the on-slab slope. \(n_T\) is
cutoff-only, before masked-domain edge loss.}\tabularnewline
\toprule\noalign{}
\begin{minipage}[b]{\linewidth}\raggedright
Freed subset
\end{minipage} & \begin{minipage}[b]{\linewidth}\raggedright
\(\Delta\)IG
\end{minipage} & \begin{minipage}[b]{\linewidth}\raggedright
Paired interval
\end{minipage} & \begin{minipage}[b]{\linewidth}\raggedright
\(K\)
\end{minipage} & \begin{minipage}[b]{\linewidth}\raggedright
\(c\)
\end{minipage} & \begin{minipage}[b]{\linewidth}\raggedright
\(p\)
\end{minipage} & \begin{minipage}[b]{\linewidth}\raggedright
\(n\)
\end{minipage} & \begin{minipage}[b]{\linewidth}\raggedright
\(n_T\)
\end{minipage} \\
\midrule\noalign{}
\endfirsthead
\toprule\noalign{}
\begin{minipage}[b]{\linewidth}\raggedright
Freed subset
\end{minipage} & \begin{minipage}[b]{\linewidth}\raggedright
\(\Delta\)IG
\end{minipage} & \begin{minipage}[b]{\linewidth}\raggedright
Paired interval
\end{minipage} & \begin{minipage}[b]{\linewidth}\raggedright
\(K\)
\end{minipage} & \begin{minipage}[b]{\linewidth}\raggedright
\(c\)
\end{minipage} & \begin{minipage}[b]{\linewidth}\raggedright
\(p\)
\end{minipage} & \begin{minipage}[b]{\linewidth}\raggedright
\(n\)
\end{minipage} & \begin{minipage}[b]{\linewidth}\raggedright
\(n_T\)
\end{minipage} \\
\midrule\noalign{}
\endhead
\bottomrule\noalign{}
\endlastfoot
Pooled reference & +0.000 & {[}+0.000, +0.000{]} & 0.218 & 0.141 & 1.206
& 0.516 & 0.441 \\
\(K\) & +0.007 & {[}−0.018, +0.034{]} & 0.188 & 0.141 & 1.206 & 0.445 &
0.381 \\
\(\alpha\) & +0.000 & {[}−0.014, +0.016{]} & 0.218 & 0.141 & 1.206 &
0.489 & 0.418 \\
\(c\), \(p\) (Omori) & +0.016 & {[}−0.014, +0.051{]} & 0.218 & 0.038 &
1.123 & 0.516 & 0.380 \\
\(d\), \(q\) (spatial) & +0.010 & {[}−0.032, +0.059{]} & 0.218 & 0.141 &
1.206 & 0.516 & 0.404 \\
\(K\), \(\alpha\) & +0.014 & {[}−0.002, +0.031{]} & 0.163 & 0.141 &
1.206 & 0.450 & 0.385 \\
\(K\), \(c\), \(p\) & +0.017 & {[}−0.013, +0.051{]} & 0.244 & 0.034 &
1.103 & 0.578 & 0.392 \\
\(c\), \(p\), \(\alpha\) & +0.024 & {[}+0.005, +0.047{]} & 0.218 & 0.033
& 1.100 & 0.596 & 0.398 \\
\(c\), \(p\), \(d\), \(q\) & +0.027 & {[}−0.025, +0.084{]} & 0.218 &
0.037 & 1.122 & 0.516 & 0.371 \\
All six (local fit) & +0.031 & {[}−0.012, +0.080{]} & 0.215 & 0.034 &
1.108 & 0.565 & 0.383 \\
\end{longtable}

A nested recency covariate multiplying the background term, with
triggering parameters fixed, has paired holdout gain +0.0120 {[}−0.0002,
+0.0221{]} nats/event and fitted coefficient 0.164. It is derived from
triggering history already available to ETAS, so a gain can reflect a
changed functional use of that history rather than new information. It
has not been tested as an increment after correcting the kernels. Its
time/space exposure quadrature remains an approximation and the
experiment is exploratory.

The rolling correction results use the law frozen before 2012 as the
reference, keeping the comparison origin-respecting for triggering
parameters. We report mean annual increments separately from mean and
median per-event increments.

\begin{longtable}[]{@{}
  >{\raggedright\arraybackslash}p{(\linewidth - 6\tabcolsep) * \real{0.2500}}
  >{\raggedright\arraybackslash}p{(\linewidth - 6\tabcolsep) * \real{0.2500}}
  >{\raggedright\arraybackslash}p{(\linewidth - 6\tabcolsep) * \real{0.2500}}
  >{\raggedright\arraybackslash}p{(\linewidth - 6\tabcolsep) * \real{0.2500}}@{}}
\caption{Rolling corrections, block length 4, paired by origin. Positive
median evidence is not suppressed when the mean interval includes zero.
The all-six row is the A−C contrast of the rolling section recomputed in
this pipeline; the two agree to \(2\times10^{-4}\) nats/year and differ
only in rounding.}\tabularnewline
\toprule\noalign{}
\begin{minipage}[b]{\linewidth}\raggedright
Correction
\end{minipage} & \begin{minipage}[b]{\linewidth}\raggedright
Mean annual increment
\end{minipage} & \begin{minipage}[b]{\linewidth}\raggedright
Mean/event
\end{minipage} & \begin{minipage}[b]{\linewidth}\raggedright
Median/event
\end{minipage} \\
\midrule\noalign{}
\endfirsthead
\toprule\noalign{}
\begin{minipage}[b]{\linewidth}\raggedright
Correction
\end{minipage} & \begin{minipage}[b]{\linewidth}\raggedright
Mean annual increment
\end{minipage} & \begin{minipage}[b]{\linewidth}\raggedright
Mean/event
\end{minipage} & \begin{minipage}[b]{\linewidth}\raggedright
Median/event
\end{minipage} \\
\midrule\noalign{}
\endhead
\bottomrule\noalign{}
\endlastfoot
\(c\), \(p\) (Omori) & −1.37 {[}−4.36, +0.64{]} & −0.009 {[}−0.053,
+0.026{]} & +0.009 {[}+0.004, +0.021{]} \\
\(K\), \(\alpha\) & −1.69 {[}−3.96, +0.35{]} & −0.022 {[}−0.061,
+0.018{]} & +0.013 {[}−0.022, +0.025{]} \\
\(K\), \(c\), \(p\) & −1.43 {[}−4.27, +0.60{]} & −0.010 {[}−0.051,
+0.024{]} & +0.014 {[}−0.001, +0.024{]} \\
\(c\), \(p\), \(\alpha\) & −1.66 {[}−5.33, +0.66{]} & −0.012 {[}−0.067,
+0.027{]} & +0.017 {[}+0.004, +0.024{]} \\
\(c\), \(p\), \(d\), \(q\) & −1.43 {[}−5.42, +1.37{]} & −0.000
{[}−0.062, +0.052{]} & +0.032 {[}−0.021, +0.056{]} \\
All six (local fit) & −2.07 {[}−6.92, +1.14{]} & −0.011 {[}−0.083,
+0.046{]} & +0.038 {[}−0.017, +0.057{]} \\
\end{longtable}

The exploratory mainshock-timing comparison stacks direct Omori
predictions against observed aftershocks of the 12 magnitude-7 and
larger sequences, including secondary triggering. The pinned pooled
kernel has a \(\chi^2\) discrepancy of 8.95 and the Omori-corrected
kernel 17.27, so the correction that scores better on the holdout
distributes early aftershocks worse in this diagnostic. The two laws
fail in opposite directions. The pooled kernel under-predicts the first
hour, 7 observed against 5.8 expected, and over-predicts hours 3 to 12,
7 against 16.1. The corrected kernel, whose faster onset is what earns
its holdout score, over-predicts throughout the first half day: 7
against 11.4 in the first hour and 7 against 15.0 in hours 3 to 12.
These are shape diagnostics: the observed counts and direct-offspring
predictions are not the same stochastic object. Matched SSN--GCMT
early-aftershock counts cannot exclude events absent from both catalogs.
Neither check establishes which correction should be deployed.

\begin{figure}
\centering
\includegraphics[width=1\linewidth,height=\textheight,keepaspectratio,alt={Correction ladder and exploratory stacked aftershock timing. The paired ladder evaluates fitted candidates; the stacked direct-kernel comparison is an exploratory shape diagnostic.}]{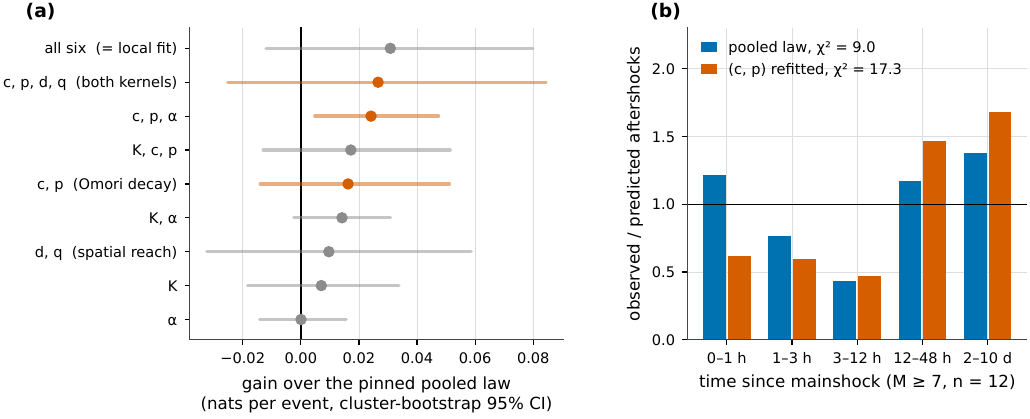}
\caption{Correction ladder and exploratory stacked aftershock timing.
The paired ladder evaluates fitted candidates; the stacked direct-kernel
comparison is an exploratory shape diagnostic.}
\end{figure}

\section{Ancillary checks: anisotropy and memory
removal}\label{ancillary-checks-anisotropy-and-memory-removal}

\subsection{Directional structure}\label{directional-structure}

The directional analysis does not fit an anisotropic ETAS alternative.
It measures the entropy deficit of parent-to-child azimuths relative to
the parent's nodal-plane strike, using local strike permutations to
retain broad geographic structure. In the near field, the reported
observed deficit is approximately 0.052 nats/pair against a local-null
mean of 0.029; one of 500 permutations reaches the observed value,
giving adjusted \(p=2/501\). Approximately 9.2\% of events have a
near-field parent under the selected definition, which scales the pair
range with the parent's rupture length (\citeproc{ref-Wells1994}{Wells
and Coppersmith 1994}). Multiplying the excess contrast by that fraction
gives a heuristic scale near 0.002 nats/event. It is neither an
estimated out-of-sample forecast gain nor an upper bound, because pair
selection, parent competition, radial dependence and background
assignment matter. Anisotropy is detected; its incremental forecast
value is untested.

\subsection{Removing pre-mainshock
history}\label{removing-pre-mainshock-history}

For each selected mainshock with Mw \(\ge7\), we find a completed 30-day
run whose daily sequence-history triggering-rate checks are below the
regional background rate. \textbf{Scoring begins after that run}, not at
its first quiet day. Thus no future quietness inside a scored window is
used to determine its start. The rate criterion sums all subsequent
regional events, is integrated over the untruncated spatial plane, and
is a pragmatic quiet-history criterion rather than inferred sequence
membership or exact domain-integrated intensity.

The full and reset forecasts differ only in whether parents before the
mainshock may trigger. A placebo removes history before a date one year
earlier. Parameters are from the pooled fit ending in 2016, including
for older sequences; so this remains a retrospective diagnostic. The
causal timing correction does not eliminate that parameter-estimation
limitation. Selected mainshocks and longer scored windows also need not
be mutually independent.

\begin{longtable}[]{@{}
  >{\raggedright\arraybackslash}p{(\linewidth - 14\tabcolsep) * \real{0.1250}}
  >{\raggedright\arraybackslash}p{(\linewidth - 14\tabcolsep) * \real{0.1250}}
  >{\raggedright\arraybackslash}p{(\linewidth - 14\tabcolsep) * \real{0.1250}}
  >{\raggedright\arraybackslash}p{(\linewidth - 14\tabcolsep) * \real{0.1250}}
  >{\raggedright\arraybackslash}p{(\linewidth - 14\tabcolsep) * \real{0.1250}}
  >{\raggedright\arraybackslash}p{(\linewidth - 14\tabcolsep) * \real{0.1250}}
  >{\raggedright\arraybackslash}p{(\linewidth - 14\tabcolsep) * \real{0.1250}}
  >{\raggedright\arraybackslash}p{(\linewidth - 14\tabcolsep) * \real{0.1250}}@{}}
\caption{Retrospective memory-removal diagnostic after a completed quiet
period, nats/event. The last two columns summarize the distribution
across sequences rather than the event-weighted mean.}\tabularnewline
\toprule\noalign{}
\begin{minipage}[b]{\linewidth}\raggedright
Horizon (days)
\end{minipage} & \begin{minipage}[b]{\linewidth}\raggedright
Sequences
\end{minipage} & \begin{minipage}[b]{\linewidth}\raggedright
Events
\end{minipage} & \begin{minipage}[b]{\linewidth}\raggedright
Reset − full
\end{minipage} & \begin{minipage}[b]{\linewidth}\raggedright
Sequence interval
\end{minipage} & \begin{minipage}[b]{\linewidth}\raggedright
Reset − placebo
\end{minipage} & \begin{minipage}[b]{\linewidth}\raggedright
Median per sequence-day
\end{minipage} & \begin{minipage}[b]{\linewidth}\raggedright
Sequences reset better
\end{minipage} \\
\midrule\noalign{}
\endfirsthead
\toprule\noalign{}
\begin{minipage}[b]{\linewidth}\raggedright
Horizon (days)
\end{minipage} & \begin{minipage}[b]{\linewidth}\raggedright
Sequences
\end{minipage} & \begin{minipage}[b]{\linewidth}\raggedright
Events
\end{minipage} & \begin{minipage}[b]{\linewidth}\raggedright
Reset − full
\end{minipage} & \begin{minipage}[b]{\linewidth}\raggedright
Sequence interval
\end{minipage} & \begin{minipage}[b]{\linewidth}\raggedright
Reset − placebo
\end{minipage} & \begin{minipage}[b]{\linewidth}\raggedright
Median per sequence-day
\end{minipage} & \begin{minipage}[b]{\linewidth}\raggedright
Sequences reset better
\end{minipage} \\
\midrule\noalign{}
\endhead
\bottomrule\noalign{}
\endlastfoot
30 & 101 & 115 & −0.0191 & {[}−0.0540, +0.0112{]} & −0.0181 {[}−0.0531,
+0.0092{]} & \(+8.7\times10^{-4}\) & 70\% \\
90 & 101 & 309 & −0.0030 & {[}−0.0178, +0.0112{]} & −0.0027 {[}−0.0157,
+0.0086{]} & \(+5.9\times10^{-4}\) & 71\% \\
365 & 101 & 1284 & −0.0033 & {[}−0.0101, +0.0026{]} & −0.0028
{[}−0.0088, +0.0021{]} & \(+1.1\times10^{-4}\) & 59\% \\
1095 & 101 & 4063 & −0.0034 & {[}−0.0066, −0.0006{]} & −0.0018
{[}−0.0040, +0.0001{]} & \(+2.2\times10^{-6}\) & 50\% \\
\end{longtable}

Windows selected using their own future quietness show an apparent
30-day benefit; that selection is circular, and such windows are
excluded. The results must be read by horizon and comparator: a negative
reset-minus-full interval means that retaining history scores better,
while an unresolved reset-minus-placebo interval does not establish a
rupture-specific effect. These tests manipulate the existing ETAS
triggering sum; they do not test whether a physical latent stress or
damage state is reset by rupture. All four horizons are reported.

The rule applied to the rolling contrasts, that median evidence is
reported alongside the mean, applies here too, and the two summaries
disagree in the same way. The event-weighted mean is negative at every
horizon, so retaining history scores better on it, yet the reset
forecast is the better one for 70\% of sequences at 30 days and for a
majority at 90 and 365 days, and the median per-sequence-day gain is
positive at all four horizons, though it decays to nothing by 1,095
days, where the reset advantage holds for almost exactly half the
sequences. The two summaries are reconciled by concentration: the mean
is dominated by a small number of event-rich sequences in which removing
history is costly, while the typical sequence is barely affected either
way. The largest single-sequence effect at 30 days is 1.11 nats/event.
We report the mean as the operative summary because a forecast service
is exposed to events rather than to sequences, and the median because a
reader comparing sequences needs a per-sequence summary. Against the
placebo, no rupture-specific effect is resolved at any horizon.

\section{Numerical checks, interpretation and
limitations}\label{numerical-checks-interpretation-and-limitations}

\begin{longtable}[]{@{}llll@{}}
\caption{Fixed-parameter SSN spatial quadrature sensitivity. Parameters
are held fixed across grids, which isolates numerical evaluation
sensitivity.}\tabularnewline
\toprule\noalign{}
Radial nodes & Angles & IG/event & Realized compensator \\
\midrule\noalign{}
\endfirsthead
\toprule\noalign{}
Radial nodes & Angles & IG/event & Realized compensator \\
\midrule\noalign{}
\endhead
\bottomrule\noalign{}
\endlastfoot
280 & 256 & 1.003252 & 121.298178 \\
560 & 512 & 1.003270 & 121.296298 \\
1120 & 1024 & 1.003265 & 121.296828 \\
\end{longtable}

\begin{longtable}[]{@{}lllll@{}}
\caption{Sharp-event counts under spatial grid refinement. Per-event S
values are supplied to show boundary sensitivity.}\tabularnewline
\toprule\noalign{}
Grid & \(S\ge0.05\) & \(S\ge0.25\) & \(S\ge1\) & \(S\ge2\) \\
\midrule\noalign{}
\endfirsthead
\toprule\noalign{}
Grid & \(S\ge0.05\) & \(S\ge0.25\) & \(S\ge1\) & \(S\ge2\) \\
\midrule\noalign{}
\endhead
\bottomrule\noalign{}
\endlastfoot
150×150 & 44 & 28 & 20 & 17 \\
300×300 & 44 & 28 & 20 & 17 \\
450×450 & 44 & 28 & 20 & 17 \\
\end{longtable}

The identifiability diagnostics referred to above are summarized here.
From 24 log-normal starting points per fit, with a dispersion of 0.6 in
log units, four of the five regional fits reach the same optimum at
every start. Central America is the exception, with two of 24 starts
ending on a degenerate ridge that drives the spatial kernel to unbounded
range or the Omori exponent to one; both are inadmissible and neither is
the reported fit. The pooled law reaches its optimum from all 12 starts,
with the top solutions agreeing to within 0.02\% in every coordinate.
Agreement across starts bounds the search; the shape of the surface is a
separate question, taken up next.

The curvature shows the shape. The Hessian of the log-likelihood is
negative definite at every reported optimum, so the observed
information, its negative, is positive definite, with smallest
eigenvalues from 0.19 to 14.94 in log-parameter units; but the condition
numbers run from 181 to 765, and the flattest direction is consistently
a combination of parameters. On the masked SSN fit it is p:-0.69 c:-0.58
K:+0.42, and on the small GCMT margins it rotates into the spatial pair.
This is the direction along which the likelihood-ratio bootstrap
displaces its recovered productivity and Omori coordinates, which is why
those shifts are read as movement along a ridge.

The profiles show where Wald summaries stop being usable. Of the 30
profiled parameters, 17 do not close within four standard errors on at
least one side, and 16 have a Wald lower limit at or below the boundary
of their admissible range, meaning a negative productivity or scale or
an Omori exponent below one. Those intervals are artifacts of local
curvature on a skewed surface. The profile is asymmetric even where it
closes: for the SSN fit \(K\) has a profile interval of {[}0.135,
0.538{]} against a Wald interval of {[}0.057, 0.373{]}, shifted upward
and wider on the high side. We report Wald standard errors as local
curvature, as in the fitted-parameter table, and use profiles wherever
an interval carries weight.

The spatial checks complement the temporal crossing refinement.
Agreement of segment totals with a compensator validates their
partitioning; the compensator's own accuracy is what the quadrature
table checks. The grid checks hold parameters fixed, so refitted kernel
estimates could still be sensitive to numerical resolution; threshold
flags close to a boundary may vary with the grid, and the reported
regime analysis is conditional on the declared production resolution.

All SSN fits and scores use the masked domain, and all transfer rows fix
Mexico's background rate. Evaluating likelihood exposure on the
enclosing rectangle instead of the mask would count spatial cells that
are excluded from observation, and it changes the SSN parameter
estimates.

The data still provide relatively few strongly informative sequence
years. Reported cluster and block intervals are approximate and
conditional on the observed units. The cluster interval's coverage is
measured under the fitted law at 89\% against a nominal 95\%, with
one-sided misses and for a process less concentrated than the observed
holdout; of the block intervals, only those on C alone and on the
frozen-local contrast D−C have measured coverage, in the rolling-origin
section, and the primary contrasts' coverage is unmeasured. Resolved
contrasts under these intervals should be described with their estimator
and assumptions, and unresolved contrasts cannot establish equivalence
or a maximum possible loss. The same holds for the exploratory
comparisons, many of which reuse one holdout. That holdout was also
reused during analysis development, and the leave-one-margin-out design,
which excludes each recipient from parameter estimation, does not undo
that reuse.

The pooled law is depth-blind. Mexican aftershock productivity can vary
by depth and rupture population (\citeproc{ref-SinghSuarez1988}{Singh
and Suárez 1988}; \citeproc{ref-Singh1985}{Singh et al. 1985};
\citeproc{ref-Pardo1995}{Pardo and Suárez 1995}), and margins have
different depth mixtures. A depth-stratified pooled alternative is a
scientifically relevant next model; it is not ruled out by insufficient
information for separate per-margin fits. This paper does not determine
whether observed regional parameter differences reflect depth
composition, physical variation or catalog differences.

The spatial kernel is isotropic, the magnitude model is simplified, and
completeness is assumed constant above an adopted threshold.
Cross-domain triggering and unobserved outside parents are omitted. The
SSN and GCMT analyses have different domains and magnitude conventions.
Synthetic conversion studies use a simpler background and do not prove
robustness to all catalog mechanisms. Background KDE estimation can
absorb spatial clustering and can move forecasts even when triggering
coefficients remain frozen.

Our operational interpretation is limited. Frozen pooling is a plausible
configuration when parameter stability and limited local data matter.
Its triggering law has been tested on held-out recipient margins, and
freezing avoids refitting those coefficients after each new sequence.
This does not show that freezing is universally better, that pooled and
local forecasts are equivalent, or that a user's cost of forecast
changes exceeds a measured annual score difference. A service policy
should state its forecast target and utility, evaluate complete forecast
changes, and monitor calibration prospectively.

\section{Conclusions}\label{conclusions}

\begin{enumerate}
\def\labelenumi{\arabic{enumi}.}
\item
  On the masked Mexican SSN holdout, ETAS gains 1.45 bits/event over a
  shared-map Poisson reference, with a cluster interval {[}−0.03,
  +2.75{]} bits. Two connected components carry 90.0\% of the signed
  total. This concentration motivates sequence-aware uncertainty
  reporting, without identifying an exact effective sample size. The
  interval's inclusion of zero is specific to the linking rule: it
  excludes zero under six of the nine rules swept, with the lower limit
  never further than 0.12 nats/event from zero. Under the fitted law the
  cluster interval covers at 89\% and the iid interval at 79\%, and the
  observed holdout is more concentrated than 99.6\% of simulated ones.
\item
  With Mexico's ETAS background rate and density fixed, six of seven
  foreign triggering sets score within 0.07 nats/event of the home fit
  and no paired interval excludes zero. When donor background rates are
  transferred along with the triggering law, the two smallest donors
  appear to transfer badly; that effect comes from the rate level.
  Pinning the recipient rate is itself not neutral, since donor laws
  differ in retained productivity, so the transfer claim rests on the
  leave-one-margin-out design, which re-estimates that rate and agrees.
\item
  The shared statistical specification is rejected by the parametric LR
  comparison, while BIC prefers the lower-dimensional pooled fit by
  184.2. The leave-one-margin-out table reports forecast performance on
  eight recipients excluded from triggering-parameter estimation. These
  results concern regularization and forecasting; they say nothing about
  a universal physical law. Against own fits restricted to the
  admissible domain the excluded pool's event-weighted paired gain is
  +0.0334 {[}+0.0002, +0.0670{]} nats/event. Its lower limit is barely
  above zero, it excludes zero under seven of nine linking rules, and
  its coverage is unmeasured; it is neither conclusive evidence of
  superiority nor a guarantee of noninferiority. None of the eight
  per-recipient paired intervals excludes zero under any of the nine
  rules.
\item
  The primary rolling contrast, A−C in mean annual gain, is −2.07
  {[}−6.92, +1.15{]} nats/year at block length 4. All contrasts are
  paired by origin; a paired mean coincides with the difference of
  per-candidate means, a paired median does not. Mean and median
  contrasts are reported separately in both annual and per-event form,
  and here they carry opposite signs. The designation of the annual mean
  as primary was made after the results existed. Freezing stabilizes
  triggering coefficients by construction; pooling changes the law being
  frozen. Matched freezing dates provide one sensitivity comparison and
  do not establish date independence in general. The coverage of the A−C
  interval has not been measured. In an auxiliary experiment under the
  pooled null, the length-4 interval on the frozen-local contrast D−C
  covered at 75.5\% against a nominal 95\%, because the frozen law's
  estimation error is shared by every origin. That warns of
  undercoverage in the primary comparisons without quantifying it, and
  measuring it would not by itself correct the intervals.
\item
  The Gutenberg--Richter slope of 0.95 used to convert the pooled law
  between thresholds was fixed by convention; at the adopted threshold
  this catalog gives 1.020 for the whole rectangle and 1.106 for the
  modeled on-slab subset. Substituting either leaves the mean annual
  verdict, the branching-ratio guard and variant E's own score and
  dispersion intact. The paired A−C median contrast excludes zero at the
  on-slab slope (1.106) and not at the convention and the
  whole-rectangle slope (1.020), so the whole-rectangle slope moves it
  without resolving it; the shrunk law's contrast behaves the same way.
  The movement is in the frozen pooled comparator, and the adapted laws
  hold still. Conclusions phrased on the median are conditional on this
  choice, and the convention should be replaced by a measured slope
  before any median-based deployment argument is made.
\item
  Local kernel and productivity corrections, threshold conversion,
  regime selection and history removal have distinct uncertainties and
  comparators. Choosing a deployed law remains an operational judgment
  until forecast-change costs and prospective calibration are measured.
\end{enumerate}

\section{Data availability and
reproducibility}\label{data-availability-and-reproducibility}

The source catalogs come from SSN, GCMT and the USGS Slab2 release cited
above. The complete analysis repository is not public with this article;
it contains data whose redistribution requires separate handling. The
derived-data supplement is public on GitHub at
\url{https://github.com/mansilla/tepeyol-paper1-supplement} and archived
on Zenodo under concept DOI
\href{https://doi.org/10.5281/zenodo.22780056}{10.5281/zenodo.22780056},
which resolves to the latest release. Its README states which per-event
tables reproduce SSN or GCMT event attributes and under which terms. It
includes per-event paired gains and cluster labels for the SSN score,
transfer test, pooled candidates, leave-one-margin-out comparison,
partial pooling, local correction ladder and covariate experiment. It
also supplies per-origin, per-replicate, per-seed and
numerical-refinement outputs, including the \(b\)-value sweep of the
re-referenced law and the rolling contrast.

The standalone \texttt{recompute\_supplement.py} requires only NumPy and
pandas and recomputes cluster, regime, sequence and moving-block
summaries from those derived tables. In particular, it verifies all
block-summary endpoints against the saved table. The manifest records
source and derived-table SHA-256 hashes, the base commit, the
modified-working-tree flag, domains, thresholds, seeds and numerical
settings. The raw-catalog fits, simulations and exposure calculations
still require their source data and analysis implementation; derived
summaries do not make those computations independently reproducible.

All score-bootstrap intervals use 20,000 percentile resamples. Ordinary
cluster contrasts use seed 0; the drop ladder uses the number of removed
clusters as seed; memory-reset contrasts use the horizon in days, with
one added for the placebo contrast and ten for gains per sequence-day.
Moving blocks use seed 0 and lengths 1--8. Count simulation uses 1,000
replicates, seed 1 and maximum magnitude 8.4. The LR bootstrap uses 200
replicates with seeds 1000--1199 and maximum magnitude 8.5. The
synthetic threshold study uses ten seeds per case and maximum magnitude
8.4. The two magnitude ceilings differ because the experiments simulate
different regions: 8.4 bounds the Mexican domain, slightly above its
largest recorded event, while the likelihood-ratio bootstrap simulates
all eight margins jointly, including those that have produced larger
earthquakes. Both ceilings are far enough above the thresholds that the
choice affects cascade sizes negligibly, and neither is used to compare
the two experiments with each other.

GCMT boxes, in longitude and latitude, are: Mexico (−107,−88; 12,22),
Central America (−88,−83; 8,16), Peru--Ecuador (−82,−68; −18,2), Chile
(−76,−66; −46,−18), Alaska--Aleutians (−170,−140; 50,62), Japan
(129,146; 30,46), Kamchatka--Kuriles (146,165; 43,57), and Sumatra--Java
(95,120; −11,6). Lower edges are inclusive and upper edges exclusive for
catalog ownership. Cascadia (−128,−120; 40,50) and New Zealand (173,180;
−43,−37) fail the training-count floor. The SSN domain is the enclosing
rectangle intersected with valid Slab2 cells.

\section{Acknowledgments}\label{acknowledgments}

The earthquake catalog was provided by the Servicio Sismológico Nacional
(SSN), Instituto de Geofísica, Universidad Nacional Autónoma de México.
We thank SSN personnel for station maintenance, data acquisition and
catalog availability. Moment tensors are from the Global CMT Project,
and slab geometry is from the USGS Slab2 release. The analysis uses
open-source scientific Python.

\section*{References}\label{references}
\addcontentsline{toc}{section}{References}

\protect\phantomsection\label{refs}

\end{document}